\documentclass[final]{agujournal2019}
\usepackage{url} 
\usepackage{lineno}
\usepackage[inline]{trackchanges} 
\usepackage{soul}

\usepackage{subfiles}
\usepackage{subcaption}
\usepackage{wrapfig}
\usepackage{amsmath}
\usepackage{commath}
\usepackage{booktabs}
\usepackage{color}
\usepackage{colortbl}
\usepackage{comment}
\usepackage{multicol}
\usepackage{enumitem}
\usepackage{ragged2e}
\justifying

\newtheorem{remark}{Remark}

\def\comments{1}
\newcommand{\Giacomo}[1]{\if\comments1\textcolor{blue}{{#1}}\fi}
\newcommand{\Jeremy}[1]{\if\comments1\textcolor{red}{{#1}}\fi}
\newcommand{\Mark}[1]{\if\comments1\textcolor{violet}{{#1}}\fi}
\newcommand{\Steven}[1]{\if\comments1\textcolor{brown}{{#1}}\fi}
\newcommand{\Darren}[1]{\if\comments1\textcolor{teal}{{#1}}\fi}
\newcommand{\Bob}[1]{\if\comments1\textcolor{magenta}{{#1}}\fi}

\draftfalse

\journalname{JAMES}

\begin{document}

\title{Storm Surge Modeling as an Application of Local Time-stepping in MPAS-Ocean}

\authors{Jeremy R. Lilly\affil{1}, Giacomo Capodaglio\affil{2}, Mark R. Petersen\affil{2}, Steven R. Brus\affil{4}, Darren Engwirda\affil{3}, Robert L. Higdon\affil{1}}

\affiliation{1}{Oregon State University, Department of Mathematics, Corvallis, OR, 97331-4605, USA}
\affiliation{2}{Computational Physics and Methods Group, Los Alamos National Laboratory, NM, 87545, USA}
\affiliation{3}{Fluid Dynamics and Solid Mechanics Group, Los Alamos National Laboratory, NM, 87545, USA}
\affiliation{4}{Mathematics and Computer Science Division, Argonne National Laboratory, IL, 60439, USA}

\correspondingauthor{Jeremy Lilly}{lillyj@oregonstate.edu}

\begin{keypoints}
    \item Storm surge modeling of Hurricane Sandy around Delaware Bay is used to demonstrate variable-resolution ocean time-stepping methods
    \item Local time-stepping schemes are up to 35\% faster, and produce solutions of comparable quality to higher-order globally-uniform schemes
\end{keypoints}

\begin{abstract}
    This paper presents the first scientific application of local time-stepping (LTS) schemes in the Model for Prediction Across Scales-Ocean (MPAS-O).
    We use LTS schemes in a single-layer, global ocean model that predicts the storm surge around the eastern coast of the United States during Hurricane Sandy.
    The variable-resolution meshes used are of unprecedentedly high resolution in MPAS-O, containing cells as small as 125 meters wide in Delaware Bay.
    It is shown that a particular, third-order LTS scheme (LTS3) produces sea-surface height (SSH) solutions that are of comparable quality to solutions produced by the classical four-stage, fourth-order Runge-Kutta method (RK4) with a uniform time step on the same meshes.
    Furthermore, LTS3 is up to 35\% faster in the best cases, showing that LTS schemes are viable for use in MPAS-O with the added benefit of substantially less computational cost.
    The results of these performance experiments inform us of the requirements for efficient mesh design for LTS schemes.
    In particular, we see that for LTS to be efficient on a given mesh, it is important to have enough cells using the coarse time-step relative to those using the fine time-step, typically at least 1:5.
\end{abstract}

\section*{Plain Language Summary}

In many applications of modern ocean models, it is useful to partition the globe into cells of different sizes, depending on the level of accuracy desired in a given region.
One uses smaller cells in regions where more accuracy is desired, and larger cells elsewhere in order to save on computational costs; such partitions are generally called variable-resolution meshes.
A well known limitation of explicit time-stepping methods, used to advance the temporal state of the model, is that that the largest step forward in time the model can take is limited by the size of the smallest cell in the mesh.
Global time-stepping schemes use a given time-step, whose size is determined by the size of the smallest cell in the mesh, everywhere on the mesh.
Local time-stepping (LTS) methods, allow us to select multiple time-steps based on the size of cells in a localized region.
Here, we use LTS schemes to model the storm surge around the eastern US coast during Hurricane Sandy in 2012. We show that a particular LTS scheme produces sea-surface height predictions of comparable quality to those produced by a state-of-the-art global time-stepping method and that LTS is to 35\% faster in the best cases.

\section{Introduction}
\label{sec:introduction}

Variable-resolution climate models have become increasingly popular in the past decade (e.g. \citeA{skamarock2012, danilov2017, korn2017}), as they offer the ability to create high-resolution regions within global domains, with fine control over the extent and transitions in grid-cell size. 
It is well known that the size of the largest time-step that can be used by an explicit time-stepping scheme is bounded above by the size of the smallest cell in the mesh according to the Courant–Friedrichs–Lewy (CFL) condition.
This restriction is of particular interest on meshes where the cell size varies greatly.
To optimize the computational cost of running a model on a variable-resolution mesh, one would like to select a small time-step on regions of high resolution (regions defined by \textit{small} cells), and a large time-step on regions of low resolution (defined by \textit{large} cells).
For simplicity, variable-resolution climate model components typically use global time-stepping schemes, where a uniform time-step is used on the entire computational domain.
As a result, one is forced to use a small time-step that is restricted by the CFL condition influenced by the smallest cell in the mesh even on large cells that would admit a larger time-step in the absence of smaller cells.
This approach results in unnecessary computational cost on low-resolution cells.
Local time-stepping (LTS) schemes can combat this performance bottle-neck by allowing the selection of different
time-steps on different regions of the mesh according to local CFL conditions instead of a single global one.

In this paper, we present the first scientific application of LTS in the framework of the Model for Prediction Across Scales-Ocean (MPAS-O) \cite{ringler2013} by modeling the storm surge caused by Hurricane Sandy on the eastern coast of the United States.
MPAS-O is a global ocean model being developed at Los Alamos National Laboratory (LANL) as a part of the Department of Energy's Energy Exascale Earth System Model (E3SM) \cite{golaz2019, petersen2019}.
MPAS-O is a multi-layer, primitive equation global ocean model that uses the TRiSK scheme for spatial discretization \cite{thuburn2009, ringler2010}, which is a staggered Arakawa C-grid method \cite{arakawa1977} defined on a Voronoi tessellation \cite{okabe2017, ju2011}.
The vertical coordinates are treated with an Arbitrary Lagrangian-Eulerian (ALE) framework, as detailed in \citeA{petersen2015}. 
A feature of MPAS-O that is of particular interest to this work is the ability to run global ocean simulations on unstructured meshes of variable resolution \cite{hoch2019}, i.e. computational cells of different sizes can be used on specific regions of the globe depending on the desired degree of spatial accuracy.
For storm surge modeling and for the accurate simulation of coastal processes in general, variable resolution meshes with high resolution on the coast and low resolution in the deep ocean are the obvious choice to maximize computational performance while at the same time maintaining the accuracy of the physical predictions \cite{mandli2014, pringle2021}.

\begin{figure}[ht]
    \centering
    \includegraphics[width=0.55\textwidth]{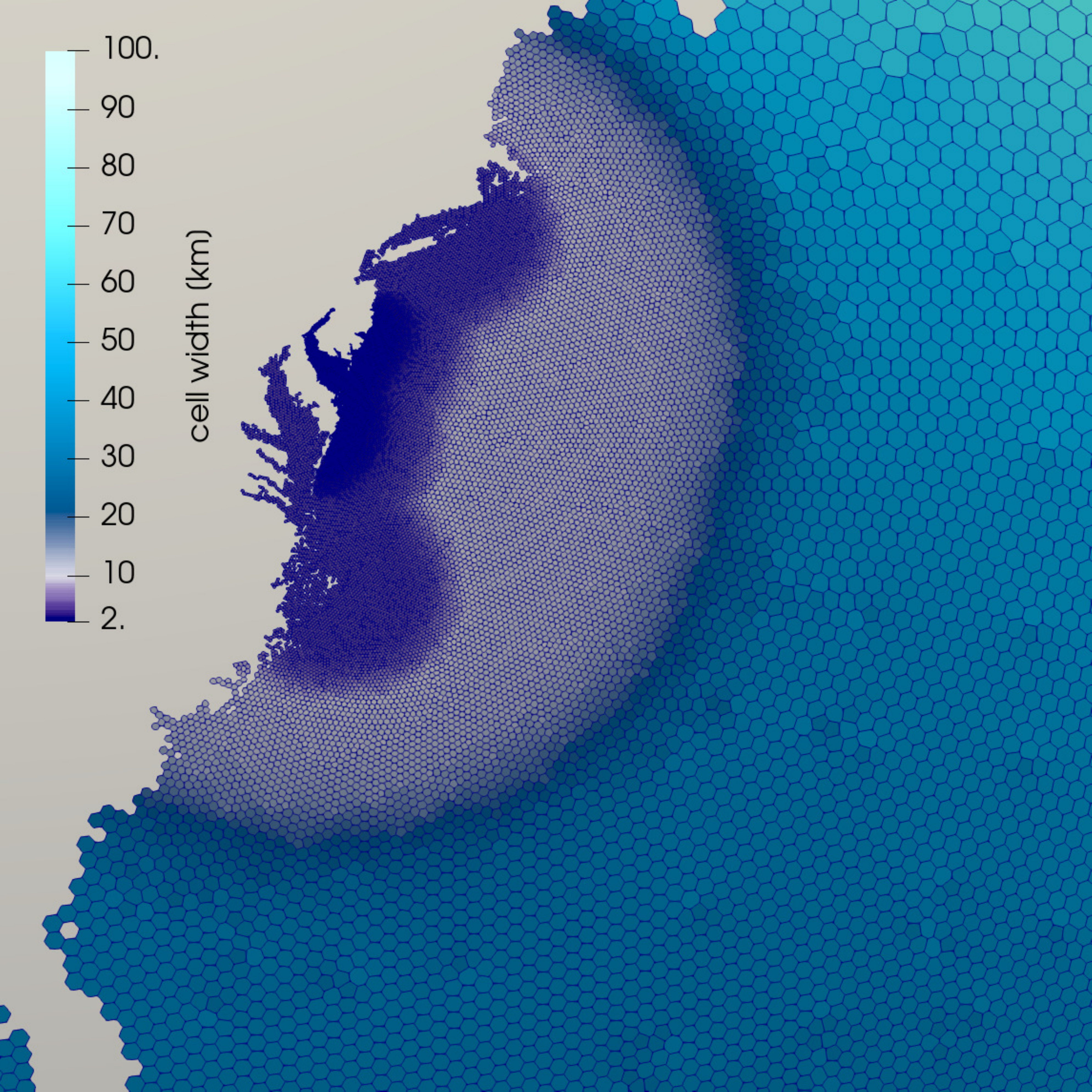}
    \caption{
        MPAS variable-resolution mesh on the East Coast of the US, with 2 kilometer resolution in Delaware Bay.
    }
    \label{fig:delaware_local}
\end{figure}

Several LTS methods have been investigated in the literature for a broad span of applications, and a comprehensive overview of these methods is beyond the scope of this work. 
Examples of such applications include solving Maxwell's equations in \citeA{montseny2008} and the wave equation in \citeA{diaz2009}, and problems in  computational aeroacoustics in \citeA{liu2010}. 
In \citeA{trahan2012} and \citeA{dawson2013}, a first-order LTS scheme has been used to solve the shallow-water equations and for storm surge modeling using a model similar to the one considered in this work.
The analysis in \citeA{dawson2013} shows that their LTS scheme produces results comparable to that of a global second order scheme, in about half the computational time.
A difference between \citeA{dawson2013} and the present work is that a discontinuous Galerkin discretization was considered in the former, whereas MPAS-O relies on the TRiSK scheme.
Moreover, our choice of LTS scheme is not the same as in \citeA{dawson2013}, but is the third order method developed in \citeA{hoang2019}.
The schemes in \citeA{hoang2019} are particularly relevant to the MPAS framework as they were developed specifically to solve the shallow water equations (SWEs) using TRiSK horizontal discretizations.
These time-stepping schemes have been implemented by the authors in the shallow water core of MPAS-O and it has been shown that they can solve the SWEs up to 70\% faster than higher-order global methods \cite{capodaglio2022}.

The focus of the current work is to explore the use of LTS in MPAS-O via a real-world application, with the goal of showing that LTS can produce virtually the same numerical solution as higher-order global methods in considerably less time.
Specifically, we use LTS to model the storm surge in the region of Delaware Bay during Hurricane Sandy and consider meshes with regions of very high resolution, such as the one in Figure \ref{fig:delaware_local}.
This particular mesh contains cells with 2 kilometer width in Delaware Bay and a global background resolution of 120 kilometers, but the meshes that will be used throughout this work have unprecedentedly high resolution near the coast, with cell widths as small as 125 meters on Delaware Bay.
We are able to run at such high resolutions by using a single-layer ocean model, as opposed to the multi-layer model, with 60 to 80 layers, that is the default for MPAS-O.
Running a single-layer significantly reduces the computational cost of the model due to fewer array elements to work on, and fewer terms in the equations, as vertical advection, diffusion, and parameterizations are off. 
The choice of a single-layer model for this study was motivated by two reasons.
First, the processes relevant to a storm surge model are essentially barotropic in nature, so a single layer is a good approximation of the physics involved.
Second, the LTS schemes developed in \citeA{hoang2019} were not developed for a layered model and as such do not take into account vertical transport between layers; work is currently underway to adapt these schemes to a layered model.

The paper is structured as follows. We begin by describing the Hurricane Sandy test case and model configuration.
We then give a brief background on the LTS schemes developed in \citeA{hoang2019} and a discussion of the inherent challenges of configuring a mesh for running with LTS. 
Next, we provide an in-depth description of a set of meshes of increasingly high resolution on which we run our Hurricane Sandy simulations.
Finally, we compare the performance in terms of CPU-time of LTS scheme of order three in \citeA{hoang2019} (LTS3) to the classical fourth-order, four-stage Runge-Kutta method (RK4) and show that LTS3 can obtain a sea-surface height solution substantially faster than RK4, with minimal differences in the quality of the prediction.


\section{Methods}
\label{sec:methods}


\subsection{Hurricane Sandy Test Case}
\label{subsec:hurricane_sandy}

The physical processes involved in the storm surge created by a hurricane are essentially barotropic in nature and can therefore be modeled by a single-layer ocean, the most important physical processes being forcing due to tidal cycles, surface wind stress, and atmospheric pressure \cite{pringle2021, mandli2014}.
Historically, MPAS-O has been used primarily for long-term (on the order of hundreds of years) climate modeling; such models require a multi-layered ocean that can resolve both barotropic and baroclinic motions.
As such, it is uncommon for MPAS-O to be run in a single layer configuration and only recently was the ability to compute tidal forcing added to the model \cite{brus2022, barton2022}.

\begin{figure}[h]
    \centering
    \includegraphics[width=0.8\textwidth]{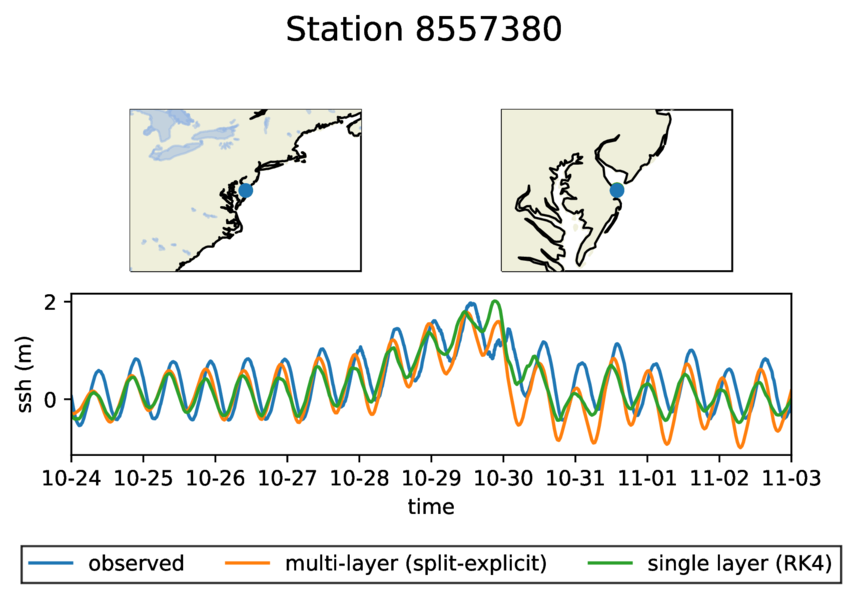}
    \caption{
        Sea-surface height (SSH) during Hurricane Sandy as predicted by a multi-layer ocean model and single-layer model considered here compared to observed data on the DelBay2km mesh (see Table \ref{tbl:meshes}).
        Split-explicit refers to the first-order, explicit, split barotropic-baroclinic time-stepping scheme that is the method of choice for a multi-layer model in MPAS-O \protect\cite{ringler2013}.
    }
    \label{fig:rk4_ssh}
\end{figure}

Figure \ref{fig:rk4_ssh} shows observed sea-surface height (SSH) at a particular tidal gauge versus time compared to the SSH predicted by MPAS-O in a multi-layer configuration and a single-layer configuration.
In Figure \ref{fig:rk4_ssh}, the multi-layered model uses a barotropic-baroclinic split-explicit time-stepping scheme \cite{higdon2005}, and the single-layer model uses RK4.
This figure shows that both the single-layer and multi-layer models include the relevant physical processes of tidal forcing, which produces semidiurnal oscillations, and the surface winds and atmospheric pressure, which produce the storm surge.

Previous work on LTS in the MPAS framework has been devoted to the shallow water equations (SWEs) \cite{capodaglio2022}, which are similar to the single-layer, barotropic model used in this work.
As a matter of fact, the present model differs from the SWEs only in having realistic bathymetry, realistic coastlines, and additional forcing terms needed to take tides, wind, atmospheric pressure, and drag exerted on the fluid by the ocean floor into account.
The affinity between the SWEs and the present model, and the need for high local resolution (generally with cells smaller than 1-5 kilometers in width) for the accurate prediction of sea-surface height in coastal regions, make storm surge modeling a good real-world test-case for LTS.
Meshes for this use case can be generated such that most of the global ocean is resolved by relatively \textit{large} cells and only areas of particular interest are highly resolved by \textit{small} cells (see Figure \ref{fig:delaware_local}).

In \citeA{mandli2014} and \citeA{pringle2021}, where mesh design for 
storm surge modeling is discussed, both works employ spatial grids that have a high resolution around coastlines while the global ocean is generally kept at a low resolution.
On such meshes, the benefits of LTS are clear; one can use a small time-step on the relatively small region(s) of high resolution, while being free to use a much larger time-step elsewhere on the globe, with consequent savings in terms of computational time.

\subsubsection{Governing Equations}
\label{subsubsec:governing_equations}

The momentum and thickness equations are given by \eqref{eqn:model}. Here the thickness equation is conservation of volume for an incompressible fluid, where the volume is normalized by the cell area, which is constant in time.  
In these equations
\( \mathbf{u} \) is the horizontal velocity,
\( t \) is the time coordinate, 
\( f \) is the Coriolis parameter, 
\( \mathbf{k} \) is the local vertical unit vector, 
\( K = \frac{\abs{\mathbf{u}}^2}{2} \) is the kinetic energy per unit mass,
\( \rho_0 \) is the (constant) fluid density,
\( p^s \) is the surface pressure, 
\( g \) is the gravitational constant, 
\( \eta \) is the sea-surface height perturbation, 
\( \eta_{\text{EQ}} \) is the sea-surface height perturbation due to equilibrium tidal forcing \cite{barton2022}, 
\( \beta \) is the self-attraction and loading coefficient \cite{barton2022}, 
\( \chi \) is the spatially varying topographic wave drag coefficient, 
\( \frac{\mathcal{C}}{H} \) is an inverse time scale,
\( H \) is the resting depth of the ocean,
\( h \) is the total ocean thickness such that \( h = H + \eta \)
\( \mathcal{C}_{\text{D}} \) is the bottom drag coefficient, 
\( \mathcal{C}_{\text{W}} \) is the wind stress coefficient,
and \( \mathbf{u}_{\text{W}} \) is the horizontal wind velocity.
\begin{equation}
    \begin{cases}
        \dfrac{\partial \mathbf{u}}{\partial t}
            + \left(\nabla \times \mathbf{u} + f \mathbf{k}\right) \times \mathbf{u}  
            =
            & - \nabla K  
              - \frac{1}{\rho_0}\nabla p^s  
              - g\nabla\left(\eta - \eta_{EQ} - \beta\eta\right) \\[7pt] 
            & - \chi \dfrac{\mathcal{C} \mathbf{u}}{H}  
              - \mathcal{C}_{\text{D}}\dfrac{\abs{\mathbf{u}}\mathbf{u}}{h}  
              + \mathcal{C}_{\text{W}}\dfrac{\abs{\mathbf{u}_{\text{W}} - \mathbf{u} }\left( \mathbf{u}_{\text{W}} - \mathbf{u} \right)}{h} \,,  
              \\[7pt]
        \dfrac{\partial h}{\partial t} + \nabla \cdot (h \mathbf{u}) = 0 \,.
    \end{cases}
    \label{eqn:model}
\end{equation}

The wind velocity \( \mathbf{u}_{\text{W}} \) is linearly interpolated from data observed at one-hour increments between 10/10/2012 and 11/03/2012. Hurricane Sandy hit the Eastern US coast on October 29, 2012. The wind and atmospheric surface pressure data were obtained from the Climate Forecast System Version 2 (CFSv2) reanalysis product \cite{saha2014}.

The model equations are spatially discretized using a C-grid type finite volume method called the TRiSK scheme wherein the fluid thickness \( h \) is stored at cell centers and the normal component of the fluid velocity \( \mathbf{u} \) is computed at cell edges.
This method has been shown to conserve the total energy of a system and robustly simulates potential vorticity \cite{thuburn2009, ringler2010}.
The current code configuration was chosen to show the performance improvements possible with LTS. Some processes, like wetting and drying cells and spatially varying bottom friction, were not included here because they are still under development, but would be expected to improve model results relative to observations.


\subsection{Time-stepping Schemes}
\label{subsec:time_stepping_schemes}

The specific LTS scheme used in this paper was originally developed for use in the MPAS framework by \citeA{hoang2019}.
This scheme is based on so-called, strong stability preserving Runge-Kutta (SSPRK) methods, which are explicit methods for solving systems of ODEs resulting from the spatial discretization of hyperbolic conservation laws.
SSPRK methods satisfy the total variation diminishing (TVD) property, which 
 means that given a sufficiently small time-step, the total variation of an SSPRK solution will not increase over time, which implies that the solution will remain stable.
For a more complete discussion of SSPRK methods, see \citeA{gottlieb1998} and  \citeA{gottlieb2001}.

\begin{figure}[h]
    \centering
    \includegraphics[width=0.6\textwidth]{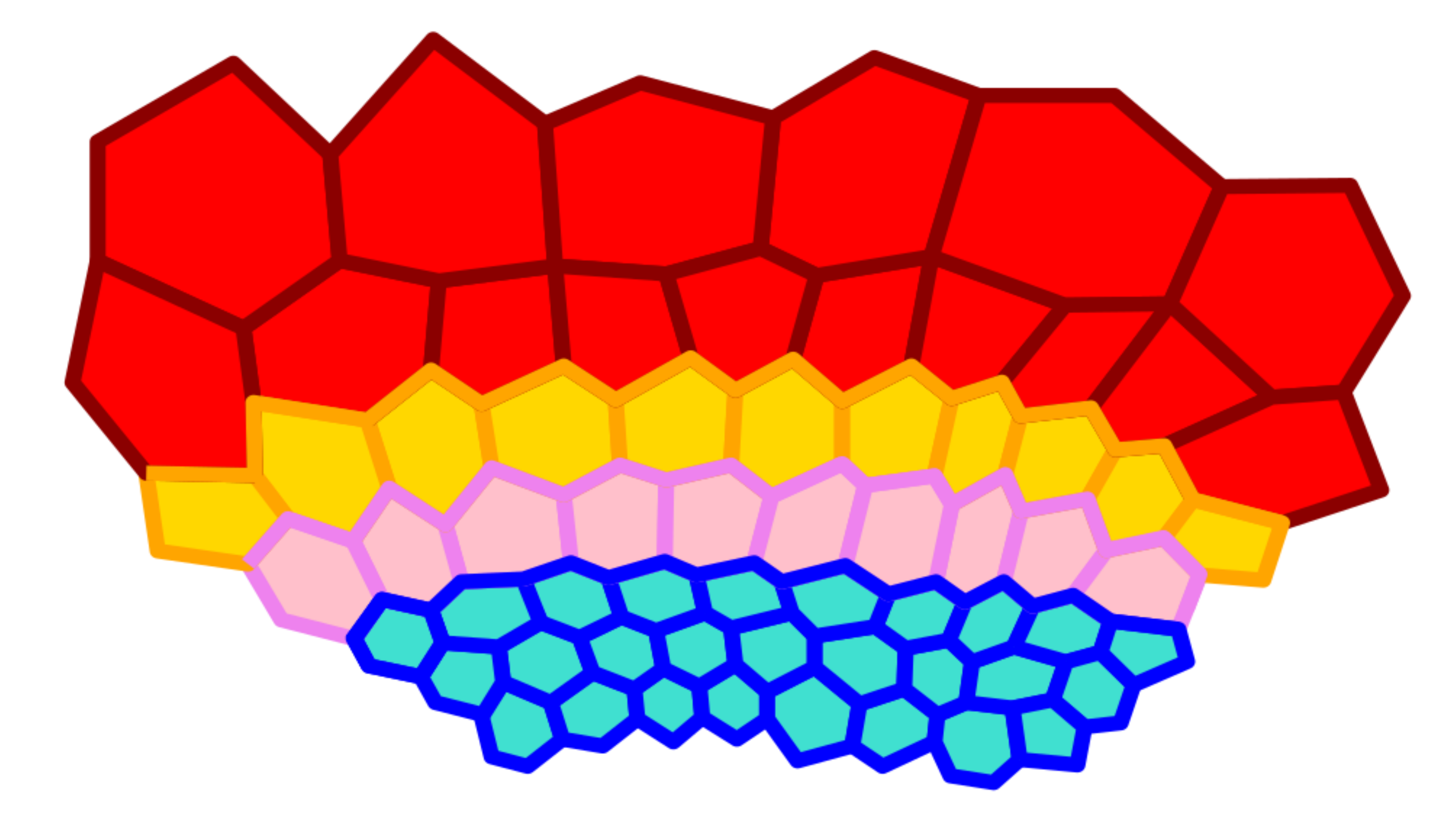}
    \caption{A mesh with cells labeled for LTS. Red cells are coarse, yellow cells are interface 2, pink cells are interface 1, and blue are fine.}
    \label{fig:lts_regions}
\end{figure}

The LTS method used here is derived from a three-stage, third-order SSPRK method (SSPRK3) and is itself third-order; we refer to this method as LTS3.
LTS3 was implemented in the MPAS framework by \citeA{capodaglio2022}.
While a full review of the LTS3 scheme is outside the scope of this paper, and can be found in \citeA{hoang2019} and \citeA{capodaglio2022}, we give a short description of how the scheme works for completeness.
Consider Figure \ref{fig:lts_regions}, which serves as a hypothetical mesh wherein there are four classes of cells.
The red cells are referred to as the coarse cells, the yellow as interface 2 cells, the pink as interface 1 cells, and the blue as fine cells.
The coarse, interface 1, and interface 2 cells all advance with the coarse time-step \( \Delta t_{\text{coarse}} \) and the fine cells advance with the fine time-step \( \Delta t_{\text{fine}} \) such that \( \Delta t_{\text{fine}} = \frac{\Delta t_{\text{coarse}}}{M} \) for some positive integer \( M \).
For practical applications, the labels \textit{fine} and \textit{coarse} denote the regions by their cell size, but the algorithm simply advances two different time steps in the two regions, so the regions could have other cell sizes---for example, identically-sized cells for testing purposes.

The LTS3 scheme proceeds as follows:
\begin{enumerate}
    \item Starting from time \( t^n \), advance the solution on the coarse, interface 1, and interface 2 cells with the first two stages of SSPRK3 using \( \Delta t_{\text{coarse}} \).
    \item Sub-cycle on the fine cells; advance to \( t^{n+1} \) by repeating all three stages of SSPRK3 \( M \) times using \( \Delta t_{\text{fine}} \).
    \begin{itemize}
        \item This requires that we know values for \( \mathbf{u} \) and \( h \) at intermediate time-levels on interface layer 1 cells. Obtain predicted values for these with an appropriate prediction step using data from time \( t^n \) and the first two stages of SSPRK3 that were obtained in step 1.
    \end{itemize}
    \item Advance to time \( t^{n+1} \) on the coarse cells with the final stage of SSPRK3.
    \item Correct the values for \( \mathbf{u} \) and \( h \) at time \( t^{n+1} \) on interface 1 and interface 2 cells by accounting for fluxes coming from the fine cells during the sub-cycling in step 2.
\end{enumerate}

\begin{remark}
    \label{remark:lts3_step1}
    \textbf{(LTS3 Step 1) }
    In order to advance with the second stage of SSPRK3 on the interface cells that border the fine region, we need the first stage SSPRK data on the three layers of fine cells adjacent to the interface region.
    Three layers are required here because the spatial discretization of the model equations mandates that a given cell's advancement depends on the three layers of adjacent cells.
    This data is not used to advance the fine cells themselves, but this fact does require that these specific fine cells, referred to as the interface-adjacent fine cells, be large enough that they admit the coarse time-step.
\end{remark}

\begin{remark}
    \label{remake:lts3_step2}
    \textbf{(LTS3 Step 2) }
    The prediction of values for \( \mathbf{u} \) and \( h \) at intermediate time-levels on interface 1 cells can be thought of as an interpolation of the existing data from \( t^n \) and the first two stages of SSPRK3. This is a substantial simplification and we refer the reader to \citeA{hoang2019} for a full description of this step.
\end{remark}

\begin{figure}[h!]
    \centering
    \begin{subfigure}{0.65\textwidth}
        \includegraphics[width=\textwidth]{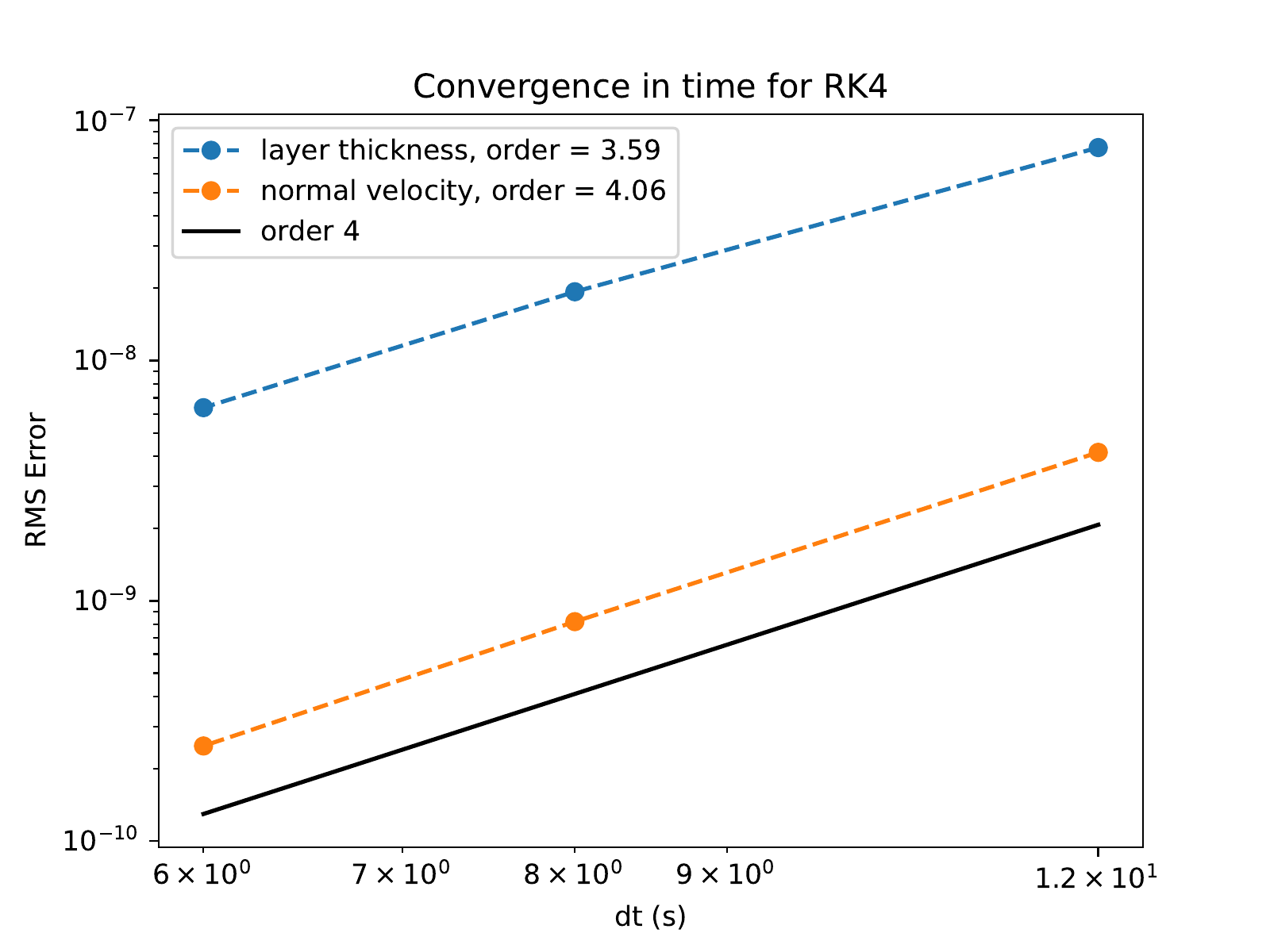}
        \caption{~}
        \label{fig:rk4_convergence}
    \end{subfigure} \\
    \begin{subfigure}{0.65\textwidth}
        \includegraphics[width=\textwidth]{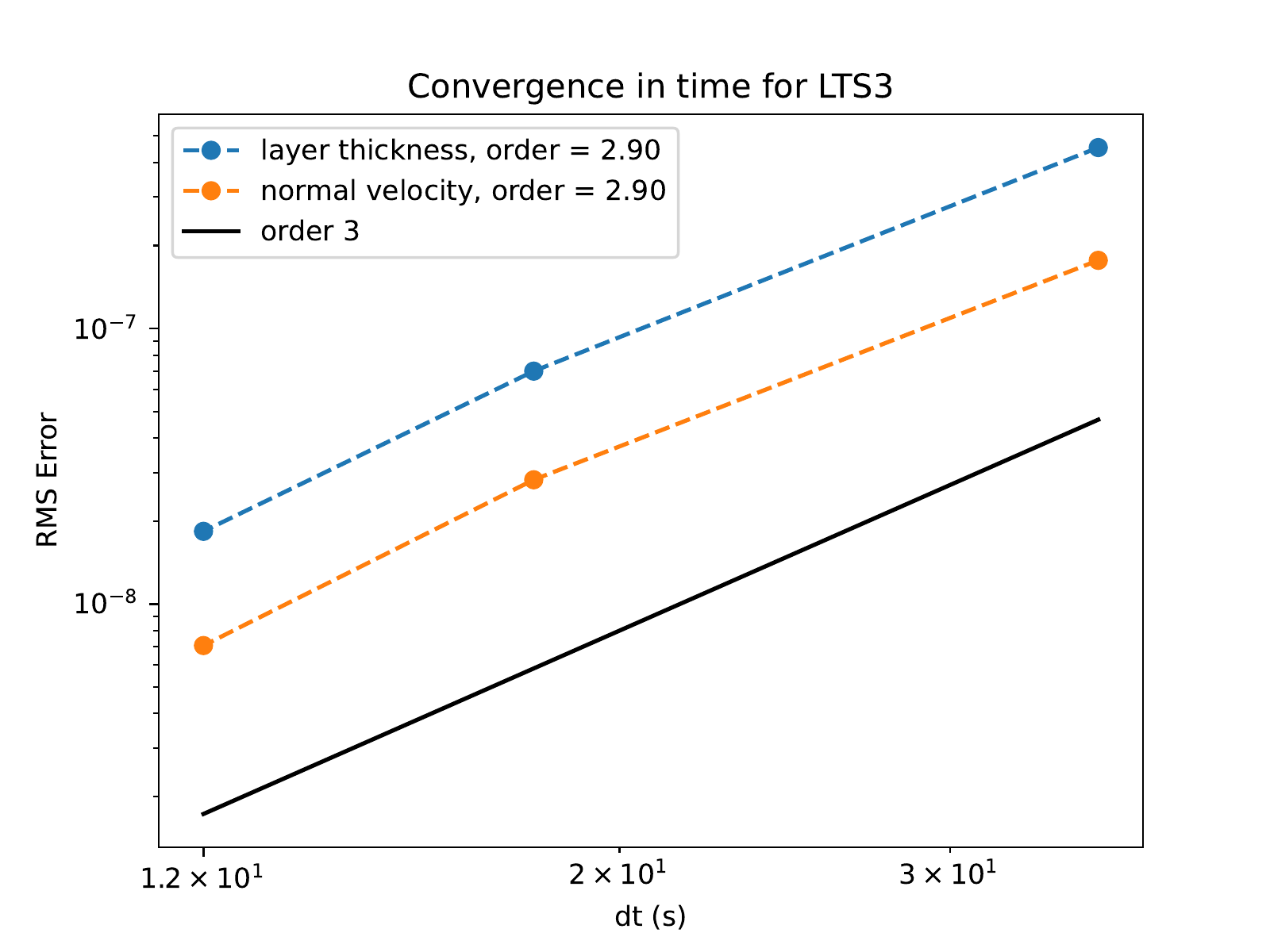}
        \caption{~}
        \label{fig:lts3_convergence}
    \end{subfigure}
    \caption{
        Temporal convergence for layer thickness \( h \) and normal velocity \( \mathbf{u} \) and  using RK4 (a) and LTS3 (b) on the discretized version of the model equations \eqref{eqn:model}. Convergence tests were conducted on the DelBay2km mesh (see Table \ref{tbl:meshes} and Figure \ref{fig:meshes_cellWidth}). For LTS3, we used \( M = 4 \). The errors were calculated against a reference solution generated by RK4 with \( \Delta t = 0.1 \). The run duration is 1 hour. The RMS error is as defined in Eq. \ref{eqn:rmse}.
    }
    \label{fig:temporal_convergence}
\end{figure}

This scheme is \( \mathcal{O}\left( (\Delta t)^3 \right) \) everywhere, including on the interface layers. 
Figures \ref{fig:rk4_convergence}  and \ref{fig:lts3_convergence} show the convergence in time for both \( \mathbf{u} \) and \( h \) using RK4 and LTS3 respectively, on the spatially discretized versions of the model equations \eqref{eqn:model}. The root-mean-square (RMS) error is defined as
\begin{equation}
    E_{\text{RMS}} = \sqrt{\frac{\sum_{i = 1}^N (s_i - m_i)^2}{N}},
    \label{eqn:rmse}
\end{equation}
where \( \{ s_i \}_{i=1}^N \) is the discrete reference solution, \( \{m_i\}_{i=1}^N \) is the discrete model solution, and \( N \) is the number of discretization points.

\begin{remark}
    \label{remark:rk4}
    One should note that RK4 uses one more Runge-Kutta stage per time-step than LTS3 and therefore has a lessened restriction on the size of the largest time-step it can use on high-resolution cells, e.g. see Table \ref{tbl:meshes} and note that on the mesh labeled DelBay2km, RK4 uses a global time-step of 30 seconds, while LTS3 uses a fine time-step of 18 seconds.
    Consider that, on the smallest cells, RK4 advances on average \( \frac{30}{4} = 7.5 \) seconds per stage while LTS3 advances on average \( \frac{18}{3} = 6 \) seconds per stage, which means that RK4 would be 25\% faster were LTS3 using a global time-step.
    This puts LTS3 at an inherent disadvantage when comparing computational performance on these meshes.
    Nevertheless, we will show that LTS3 is still capable of considerably outperforming RK4 in almost all of our case studies.
    We have chosen to use RK4 as our point of comparison rather than a third order method because RK4 is the state-of-the-art scheme in MPAS-O for a single-layer model.
\end{remark}


\subsection{Meshes and LTS Configuration}
\label{subsec:meshes}

Here we consider five different meshes; the relevant parameters for each are given in Table \ref{tbl:meshes}.
Of particular interest is the physical distribution of regions of low and high resolutions.
Each mesh is divided into five regions, each of which is populated by cells of a different size.
These regions are Delaware Bay, the area around the Delaware coast, the area around the eastern coast of the US, the western Atlantic ocean, and the rest of the globe.
In the \textit{Resolutions} section of Table \ref{tbl:meshes}, we give the cell width in kilometers for each of these five regions, for all meshes considered.
In particular, note that the highest resolutions in these meshes range from 2 kilometers down to 125 meters and the ratio of the lowest to highest resolution ranges from 60 to 240.
It should be noted that even our lowest resolution mesh using 2 kilometer cells in Delaware Bay is of extremely high resolution for MPAS-O.
Simulations with these ultra-high-resolution meshes are possible because the single-layer configuration is much faster than standard MPAS-O climate domains, which use 60 to 100 layers, vertical advection and diffusion, and additional parameterizations.

\begin{figure}[h!]
    \centering
    \begin{tabular}{lcc}
        & \multicolumn{2}{c}{\includegraphics[width=0.9\textwidth]{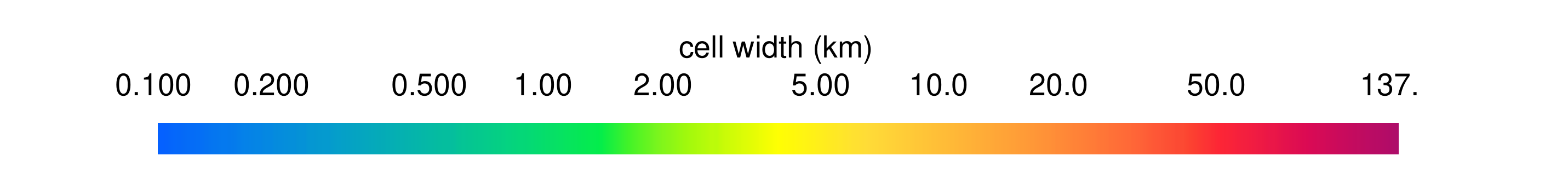}} \medskip \\
        & Atlantic Ocean & Eastern US Coast \smallskip \\
        \rotatebox{90}{\hspace{0.2cm} DelBay2km}
            & \includegraphics[width=0.45\textwidth]{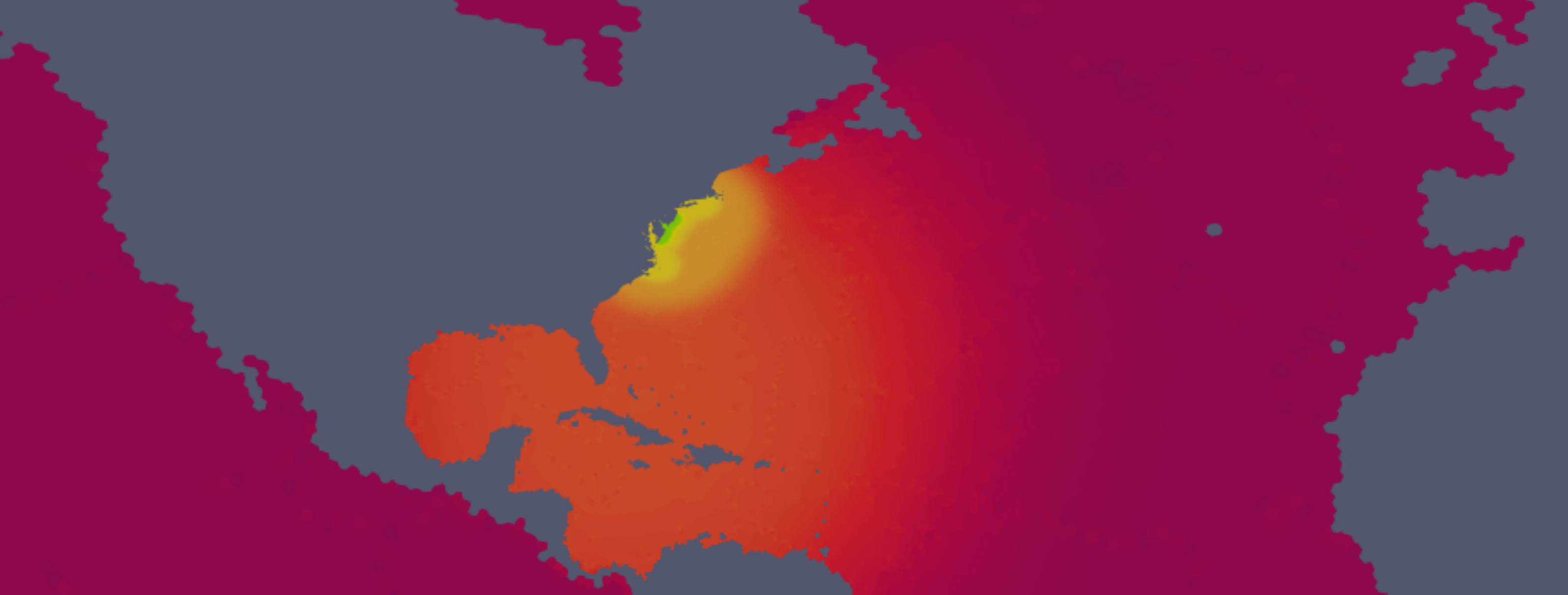}
            & \includegraphics[width=0.45\textwidth]{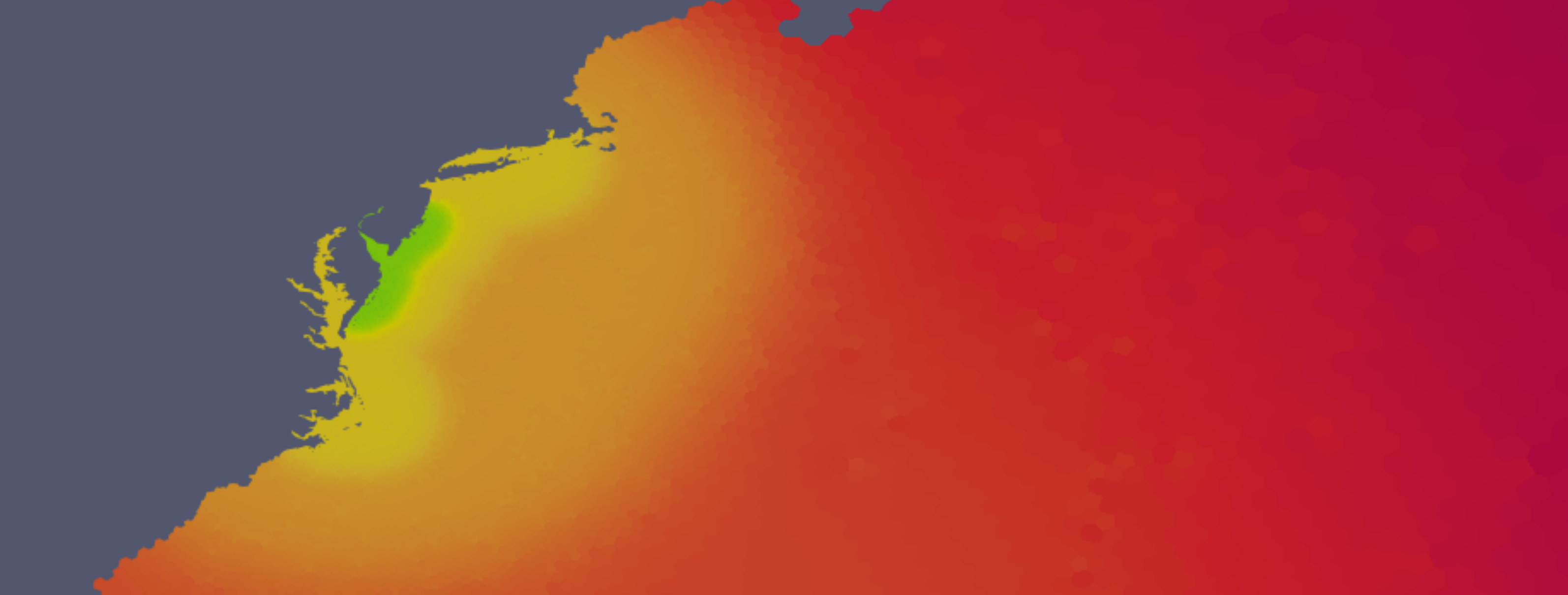} \medskip \\
        \rotatebox{90}{\hspace{0.2cm} DelBay1km}
            & \includegraphics[width=0.45\textwidth]{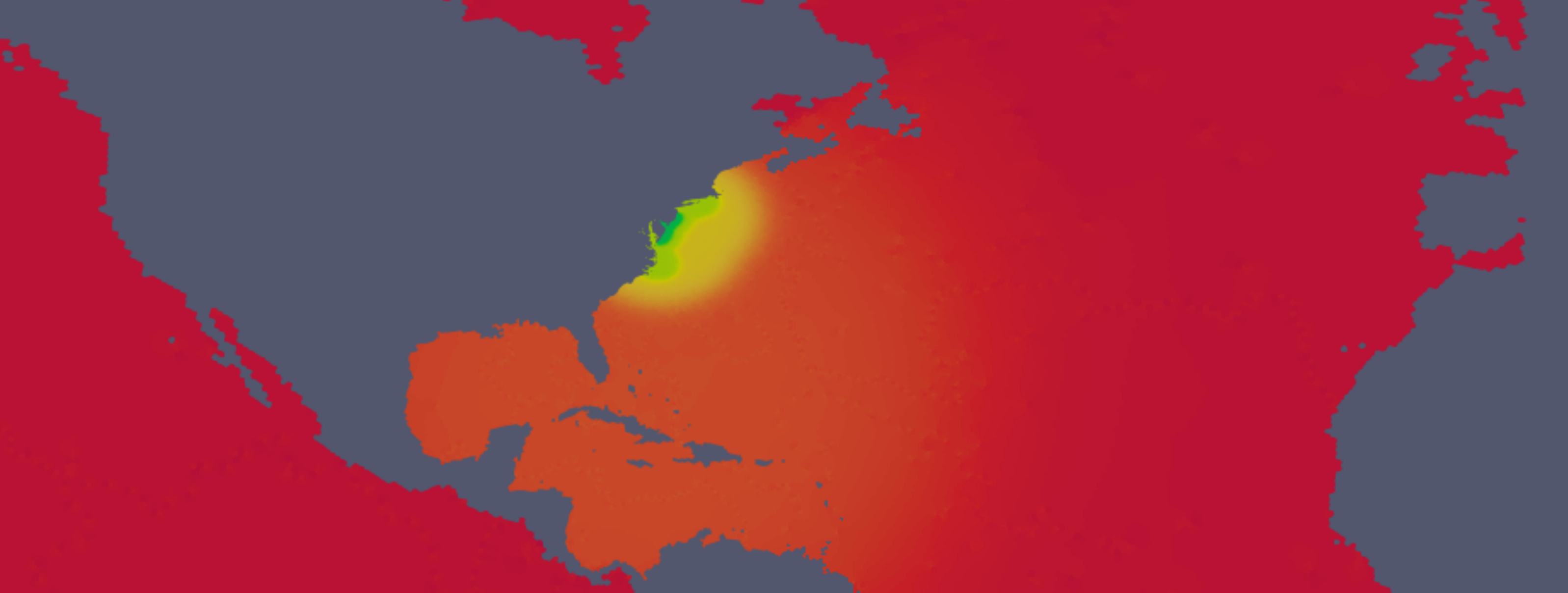}
            & \includegraphics[width=0.45\textwidth]{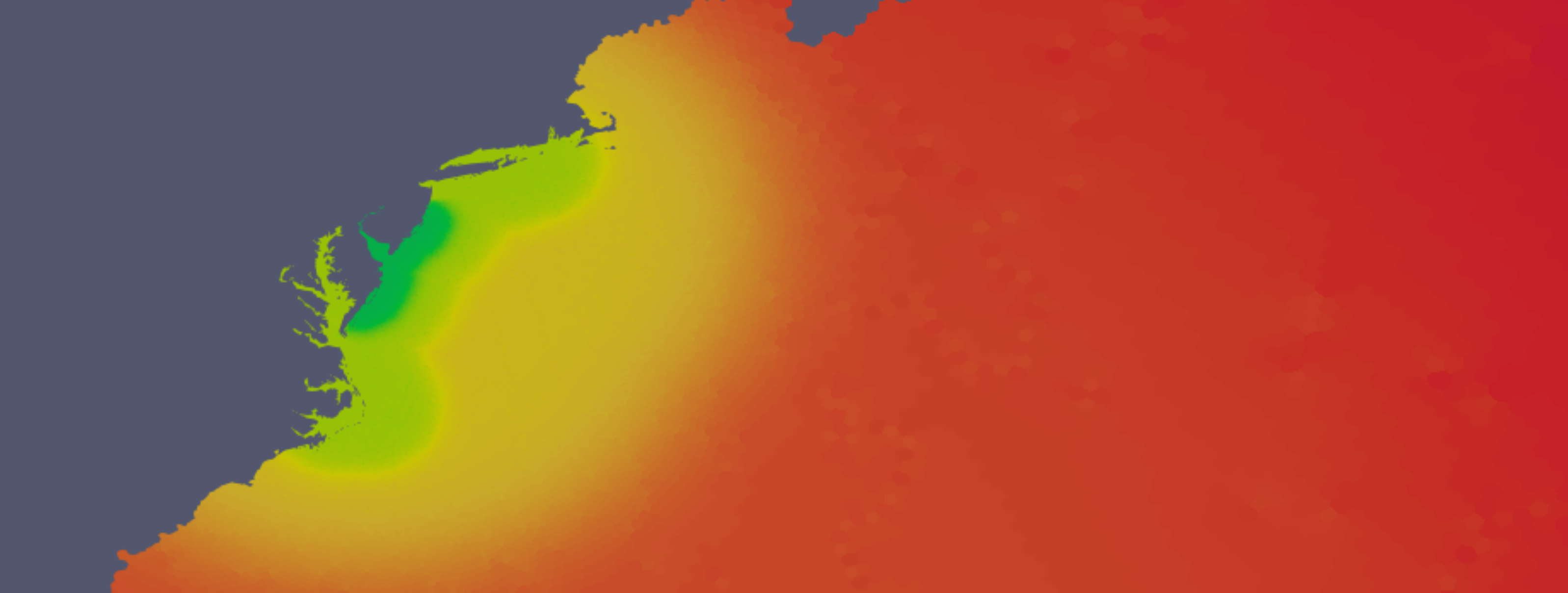} \medskip \\
        \rotatebox{90}{\hspace{0.1cm} DelBay500m}
            & \includegraphics[width=0.45\textwidth]{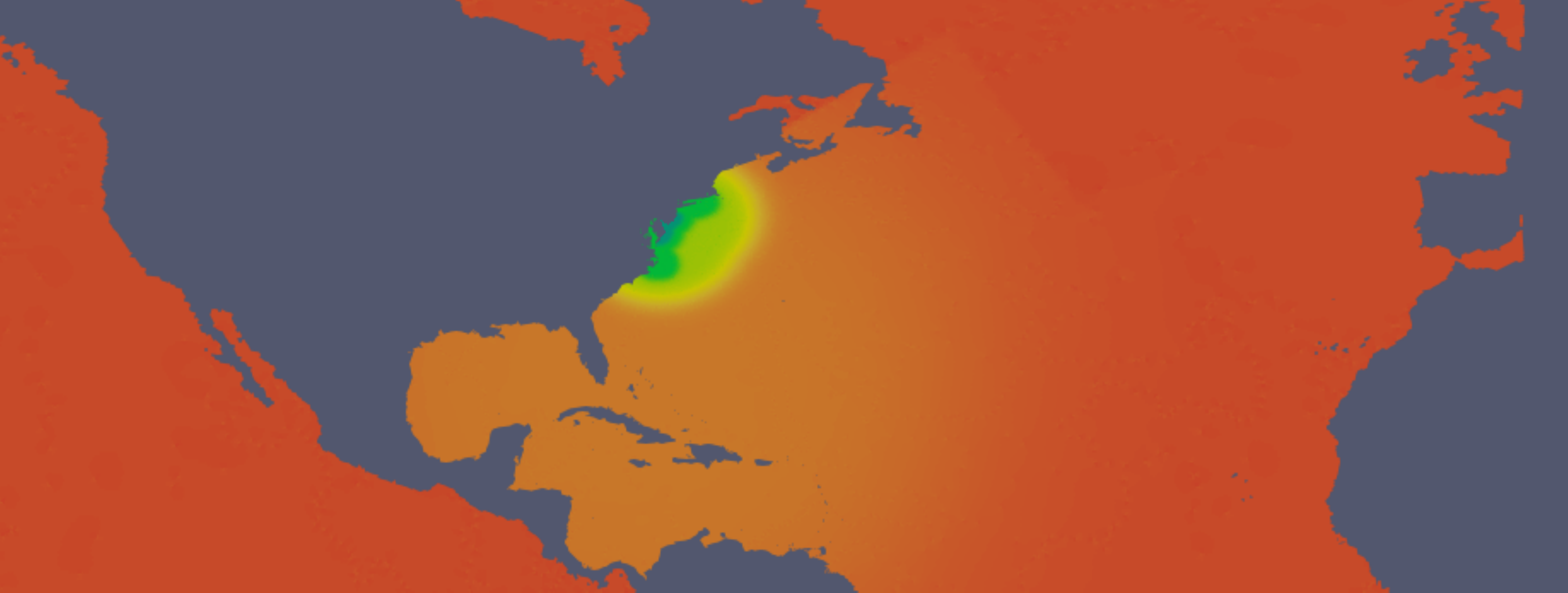}
            & \includegraphics[width=0.45\textwidth]{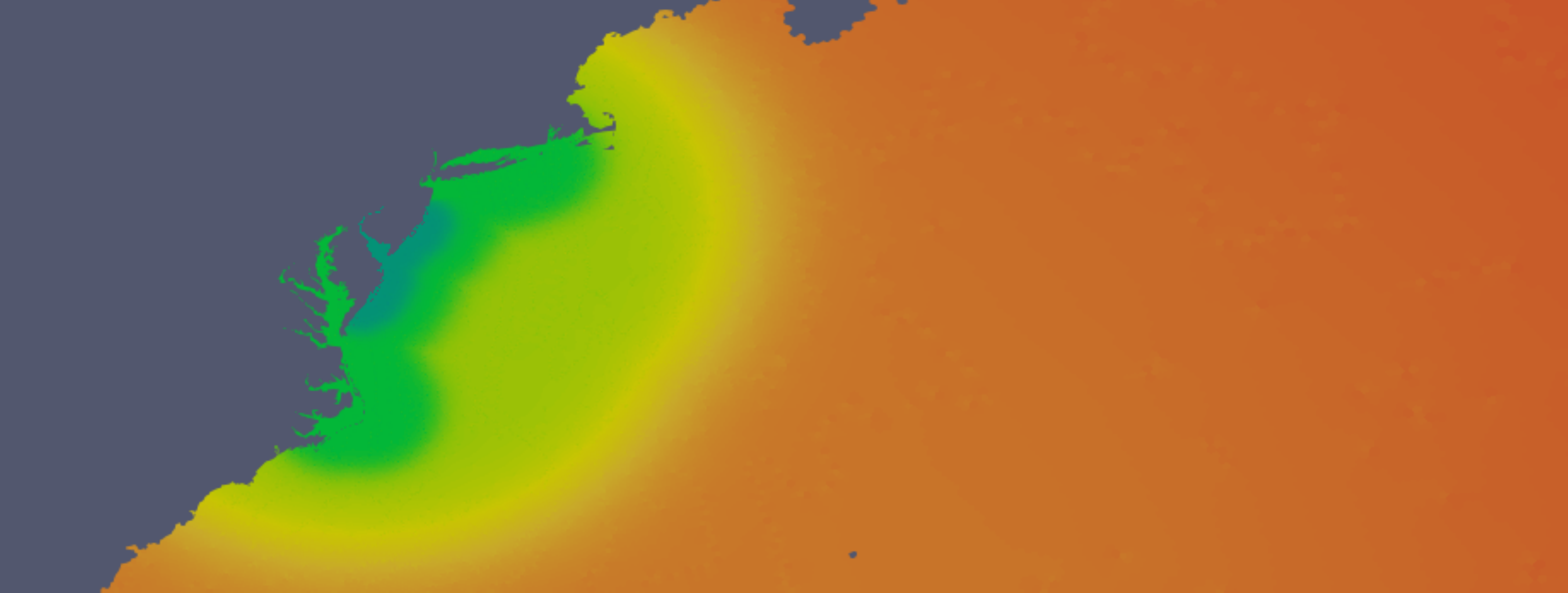} \medskip \\
        \rotatebox{90}{\hspace{0.1cm} DelBay250m}
            & \includegraphics[width=0.45\textwidth]{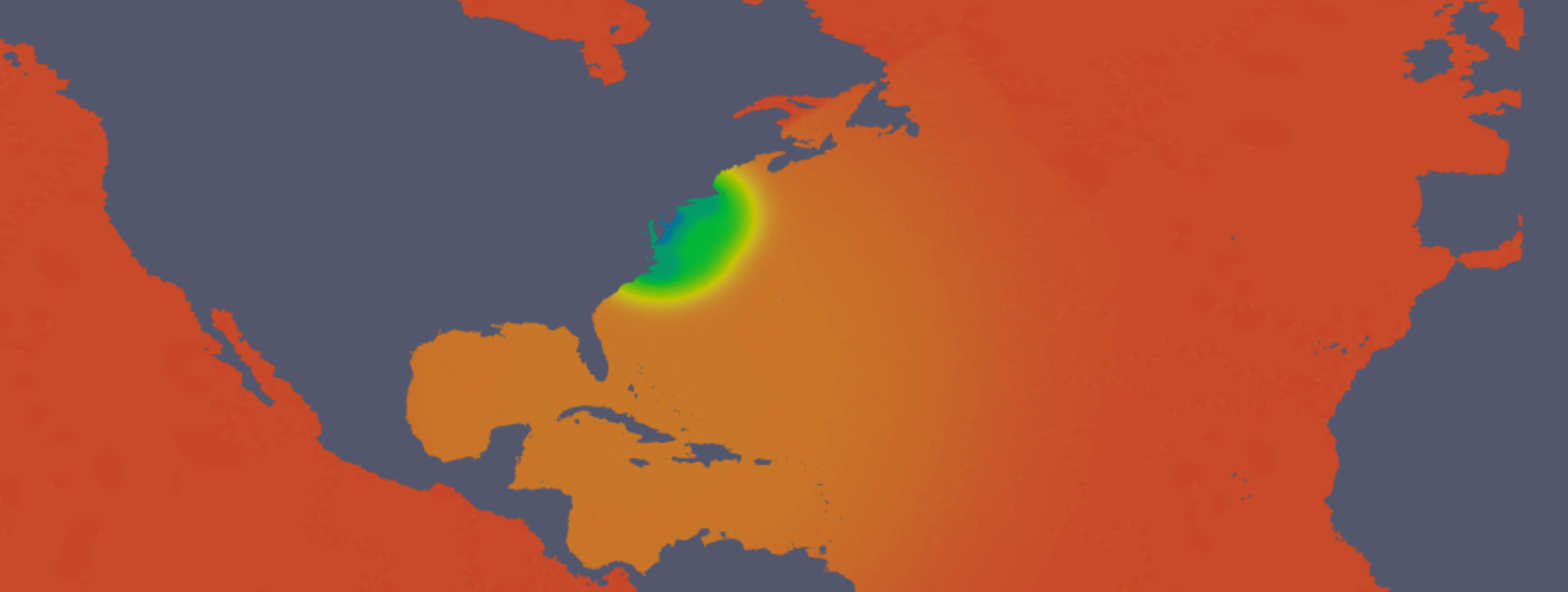}
            & \includegraphics[width=0.45\textwidth]{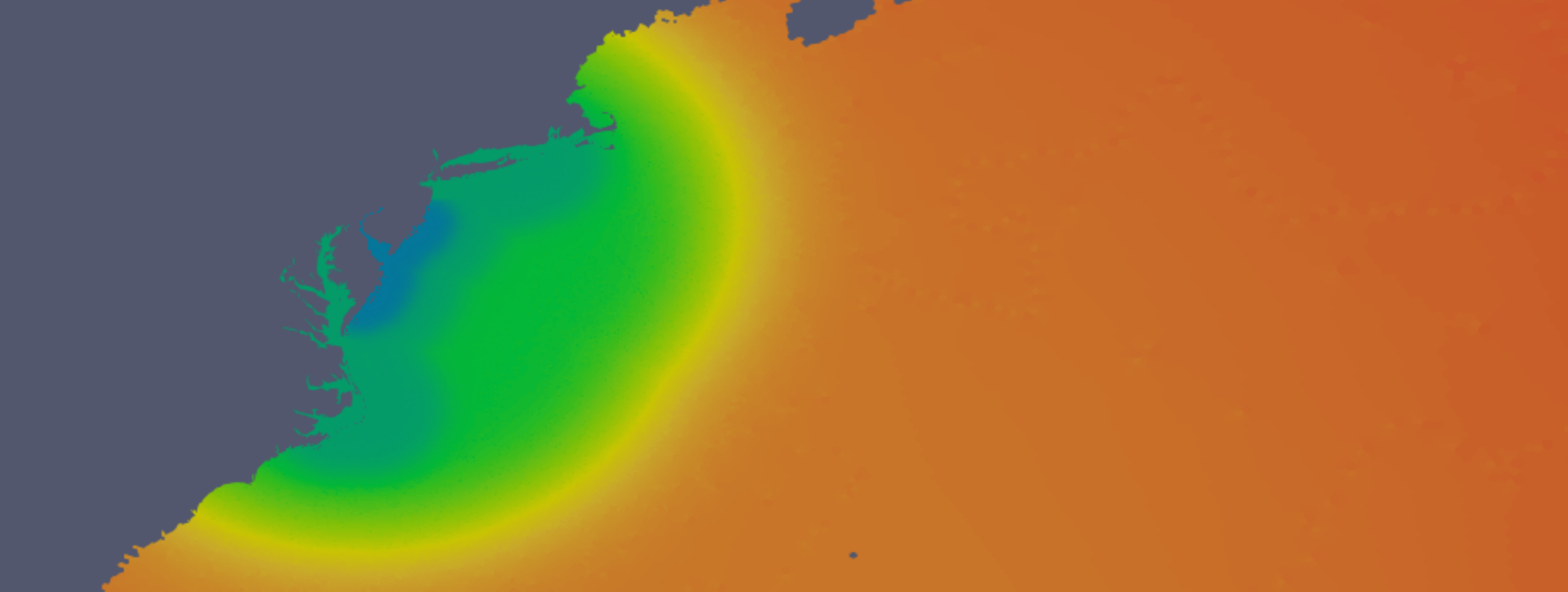} \medskip \\
        \rotatebox{90}{\hspace{0.1cm} DelBay125m}
            & \includegraphics[width=0.45\textwidth]{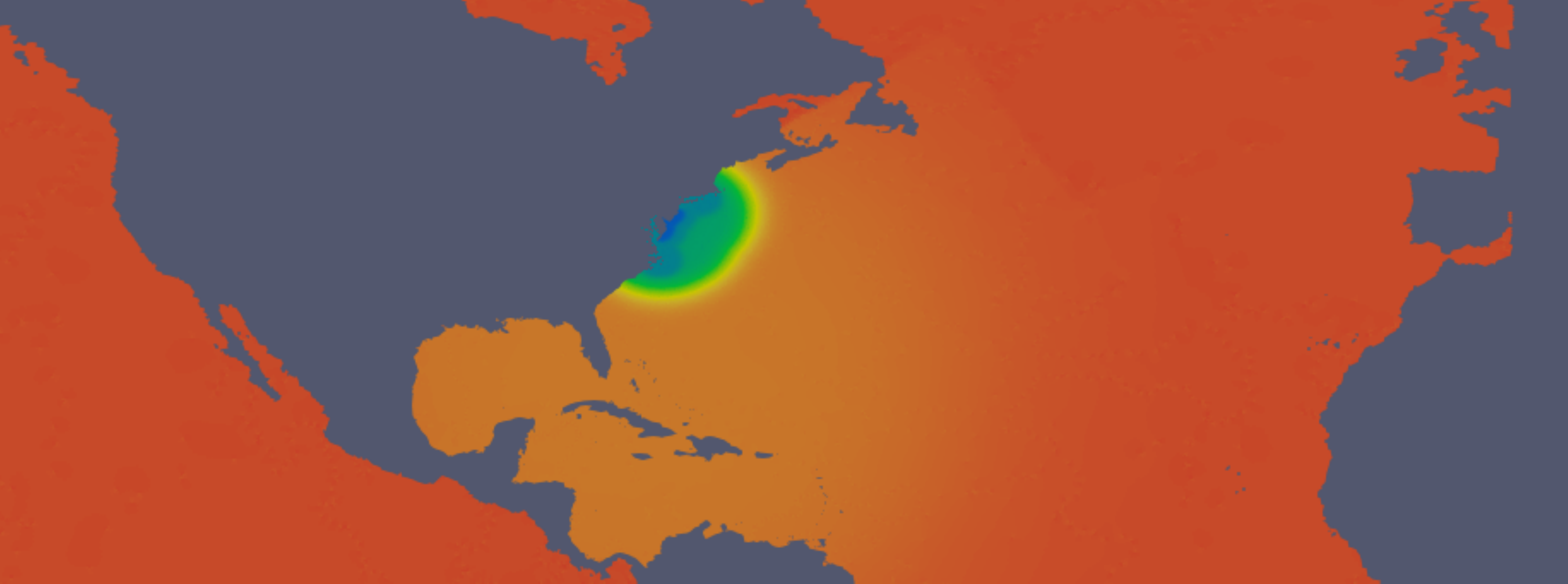}
            & \includegraphics[width=0.45\textwidth]{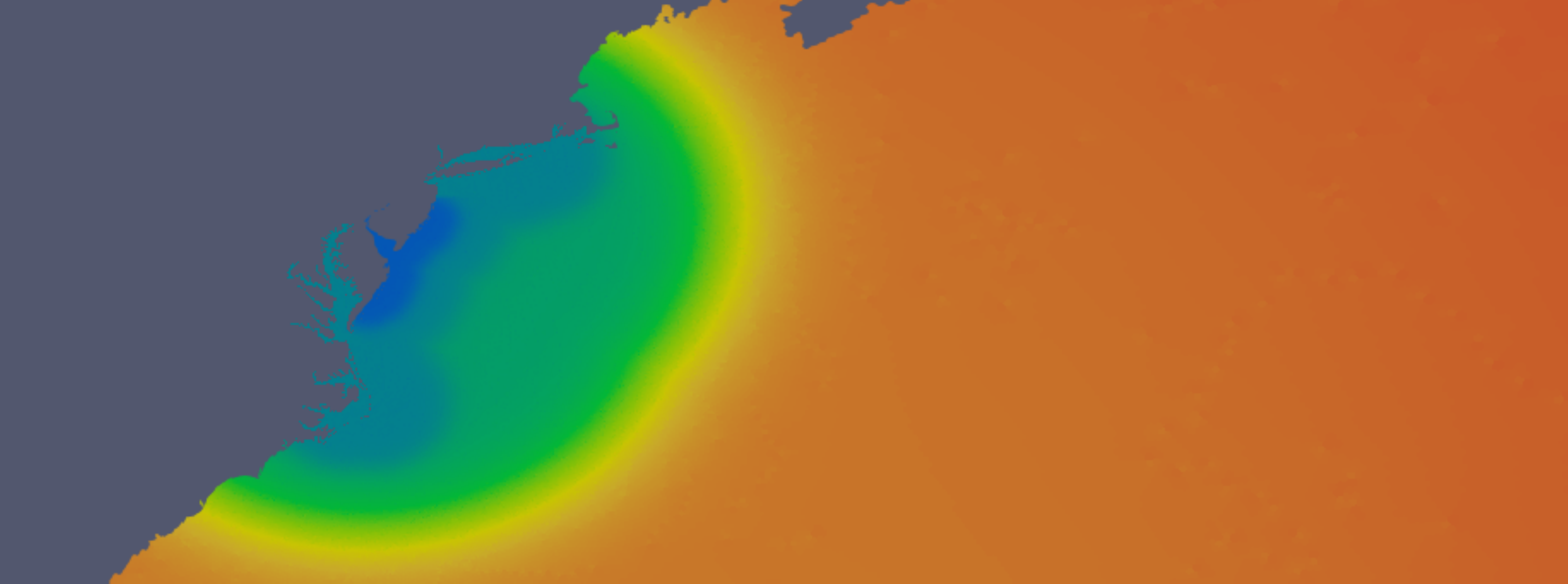} \\
    \end{tabular}
    
    \caption{
        Global cell width in kilometers for each mesh described in Table \ref{tbl:meshes}.
        Each subplot shares the same color scale, which is in log-space, so differences in high-resolution areas can be easily distinguished.
    }
    \label{fig:meshes_cellWidth}
\end{figure}

All meshes were created with JIGSAW \cite{engwirda2017}, a library that can quickly generate high-quality variable-resolution meshes.
Global meshes are designed by specifying the cell width distribution across the sphere (Figure \ref{fig:meshes_cellWidth}), which is then passed into JIGSAW.
The five regions of the mesh were used as a straightforward means of providing regional refinement in the Delaware Bay estuary within a global mesh.
The Delaware Bay region is the highest level of refinement, while the Delaware coast, and eastern US coast regions provide a smooth transition to the coarser global resolution.
The western Atlantic Ocean region uses a higher level of refinement than the global ocean to better resolve the hurricane wind and pressure fields, while the rest of the Earth has the coarsest resolution to reduce the total number of cells in the mesh.

\begin{table}[ht]
    \centering
    \resizebox{\columnwidth}{!}{%
        \begin{tabular}{lrrrrr}
            ~ & \textbf{DelBay2km} & \textbf{DelBay1km} & \textbf{DelBay500m} & \textbf{DelBay250m} & \textbf{DelBay125m} \\ \toprule
            \textbf{Resolutions:} Grid cell width (km)  & ~ & ~ & ~ & ~ & ~ \\ \midrule
            Global background & 120 & 60 & 30 & 30 & 30 \\ 
            Western Atlantic & 30 & 30 & 15 & 15 & 15  \\ 
            Eastern US coast & 10 & 5 & 2.5 & 1.25 & 0.625 \\ 
            Delaware coast & 5 & 2.5 & 1.25 & 0.625 & 0.3125  \\ 
            Delaware Bay & 2 & 1 & 0.5 & 0.25 & 0.125  \\ \toprule
            \textbf{Mesh Parameters} & ~ & ~ & ~ & ~ & ~  \\ \midrule
            Number of cells & 58,240 & 198,776 & 794,172 & 1,591,416 & 4,617,565  \\
            Number of MPI ranks & 2 & 8 & 32 & 64 & 178  \\ 
            Thousands of cells per MPI rank & 29 & 25 & 25 & 25 & 26  \\ \toprule
            \textbf{LTS Parameters} & ~ & ~ & ~ & ~ & ~  \\
            \textit{Eastern US Coast (EC) Fine Region} & ~ & ~ & ~ & ~ & ~  \\ \midrule
            Number of interface layers & 2 & 5 & 10 & 20 & 50   \\ 
            Count ratio  & 1.88 & 1.71 & 1.71 & 0.46 & 0.12  \\
            Resolution ratio & 15 & 30 & 30 & 60 & 120  \\
            \( \Delta t_{\text{RK4}} \) (s) & 30 & 15 & 7.5 & 3.75 & 1.875  \\ 
            \( \Delta t_{\text{fine}} \) (s) & 18 & 8 & 4 & 2 & 1  \\ 
            \( \Delta t_{\text{coarse}} \) (s) & 72 & 72 & 24 & 24 & 24  \\
            \( M \) & 4 & 9 & 6 & 12 & 24  \\

            \textit{Western Atlantic (WA) Fine Region} & ~ & ~ & ~ & ~ & ~   \\ \midrule
            Number of interface layers & 2 & 5 & 10 & 20 & 50  \\ 
            Count ratio  & 0.92 & 1.28 & 1.28 & 0.39 & 0.11  \\ 
            Resolution ratio & 60 & 60 & 60 & 120 & 240  \\ 
            \( \Delta t_{\text{RK4}} \) (s) & 30 & 15 & 7.5 & 3.75 & 1.875  \\ 
            \( \Delta t_{\text{fine}} \) (s) & 18 & 8 & 4 & 2 & 1  \\ 
            \( \Delta t_{\text{coarse}} \) (s) & 306 & 152 & 80 & 80 & 80  \\
            \( M \) & 17 & 19 & 20 & 40 & 80 \\ \bottomrule
        \end{tabular}
    }
    \caption{
        Relevant parameters for each mesh and LTS configuration used in performance experiments.
    }
    \label{tbl:meshes}
\end{table}

\subsubsection{Choice of Time-steps}
\label{subsubsec:time_steps}

The time-steps used on each mesh and each LTS configuration for performance tests are given in Table \ref{tbl:meshes}.
These time-steps were obtained experimentally and are the largest values that are admissible to guarantee stability of the model.

Because there are two different time-steps for the LTS scheme and these must differ by a factor of \( M \) such that \( \Delta t_{\text{fine}} = \frac{\Delta t_{\text{coarse}}}{M} \), they can be obtained in different orders; one could find the largest time-step admittable on the coarse region of the mesh then find the \textit{smallest} value of \( M \) that gave an admittable fine time-step, or first find the largest admittable fine time-step then the \textit{largest} value of \( M \) that gives an admittable coarse time-step.
This has the result of the time-step in one region or the other not technically being maximal.

There are advantages to both methods that depend on the mesh itself. For instance, on a mesh where there are sufficiently more coarse cells than fine cells, one might wish to first maximize the time-step in the coarse region and take a small penalty to the size of the fine time-step.
In this paper however, we have opted to find admittable time-steps by first maximizing the fine time-step, then by finding the largest value of \( M \) that gives a valid coarse time-step with the intention on minimizing the work done in the fine region.

\subsubsection{LTS Mesh Parameters}
\label{subsubsec:lts_mesh_parameters}

When designing a mesh for use with LTS there are two parameters of particular interest, which we refer to as the \textit{count ratio} and the \textit{resolution ratio}.
Both of these parameters depend on the number of cells where the fine time-step is used, referred to as the fine cells, and the number of cells where the coarse time-step is used, referred to as the coarse cells.
Note that, in this discussion, the label of fine or coarse does not necessarily reference the size of a cell, but rather which time-step is used to advance it.
The region made up of fine cells is called the fine region while the rest of the globe is called the coarse region.

The count ratio is the ratio of the number of cells in the coarse region to the number of cells in the fine region, i.e.
\[ \text{count ratio} = \frac{\text{number of coarse cells}}{\text{number of fine cells}}. \]
The resolution ratio is the ratio of the cell width of the coarse cells to the cell width of the fine cells, i.e.
\[ \text{resolution ratio} = \frac{\text{cell width of coarse cells}}{\text{cell width of fine cells}}. \]
In the case where there are cells of multiple resolutions in either region, as is the case in our meshes, we consider the smallest value of cell width in a given region as it is the smallest cell that restricts the size of the time-step admittable in that region.
For example, in DelBay2km with the EC configuration the smallest fine cells have a cell width of 2 kilometers and the smallest coarse cells have a cell width of 30 kilometers so we would say that the resolution ratio was 15.

These parameters are the primary drivers behind the performance of LTS.
The higher the resolution ratio is, the larger the coarse time-step can be, allowing LTS to take significantly less time-steps on coarse cells compared to a global method, which must use a small time-step everywhere.
Similarly, the higher the count ratio is, the more coarse cells there are in a mesh where LTS has to do less work than a global method.

\begin{figure}[h!]
    \centering
    \begin{subfigure}{0.8\textwidth}
        \includegraphics[width=\textwidth]{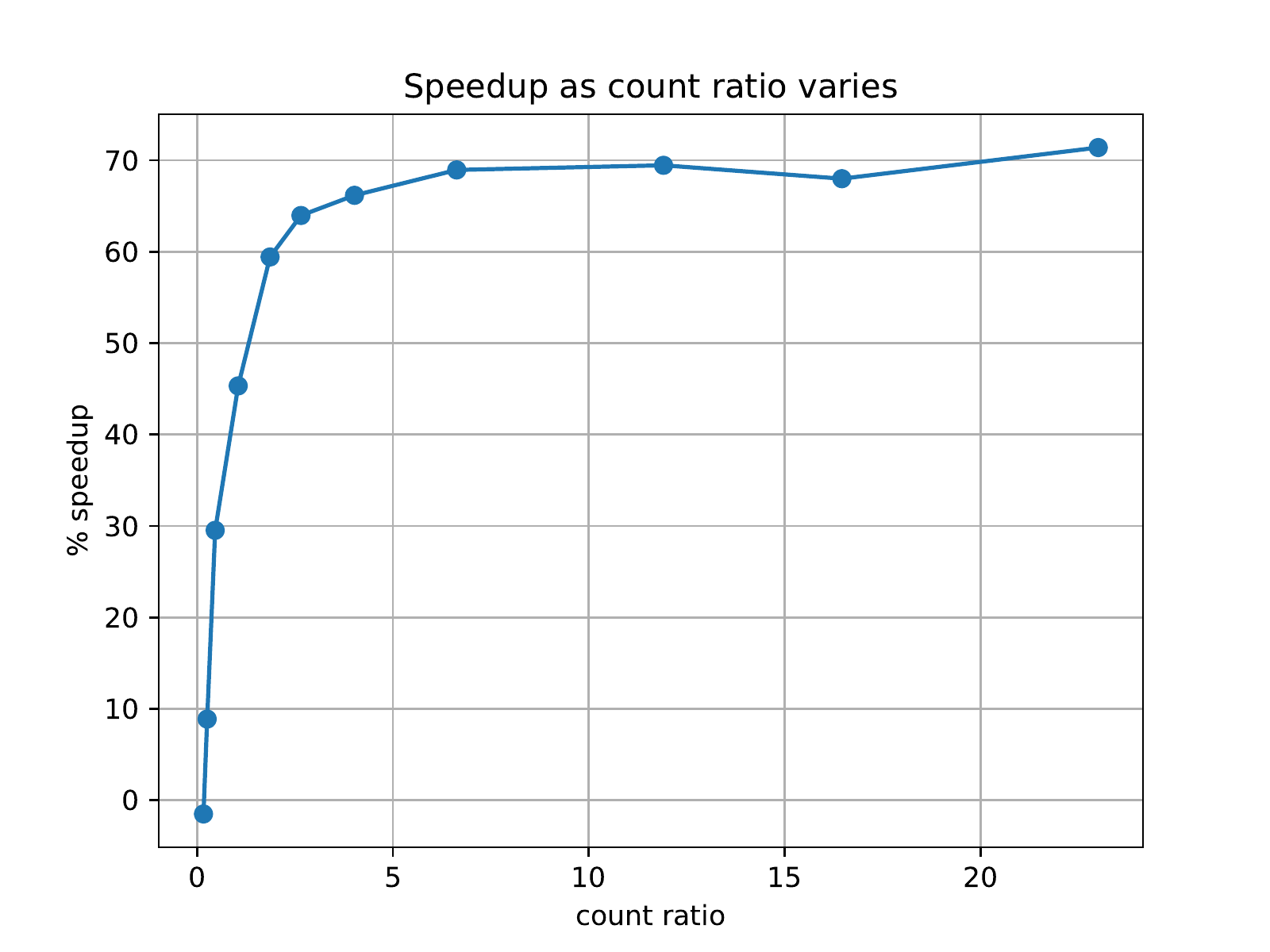}
        \caption{~}
        \label{fig:count_ratio}
    \end{subfigure} \\
    \begin{subfigure}{0.8\textwidth}
        \includegraphics[width=\textwidth]{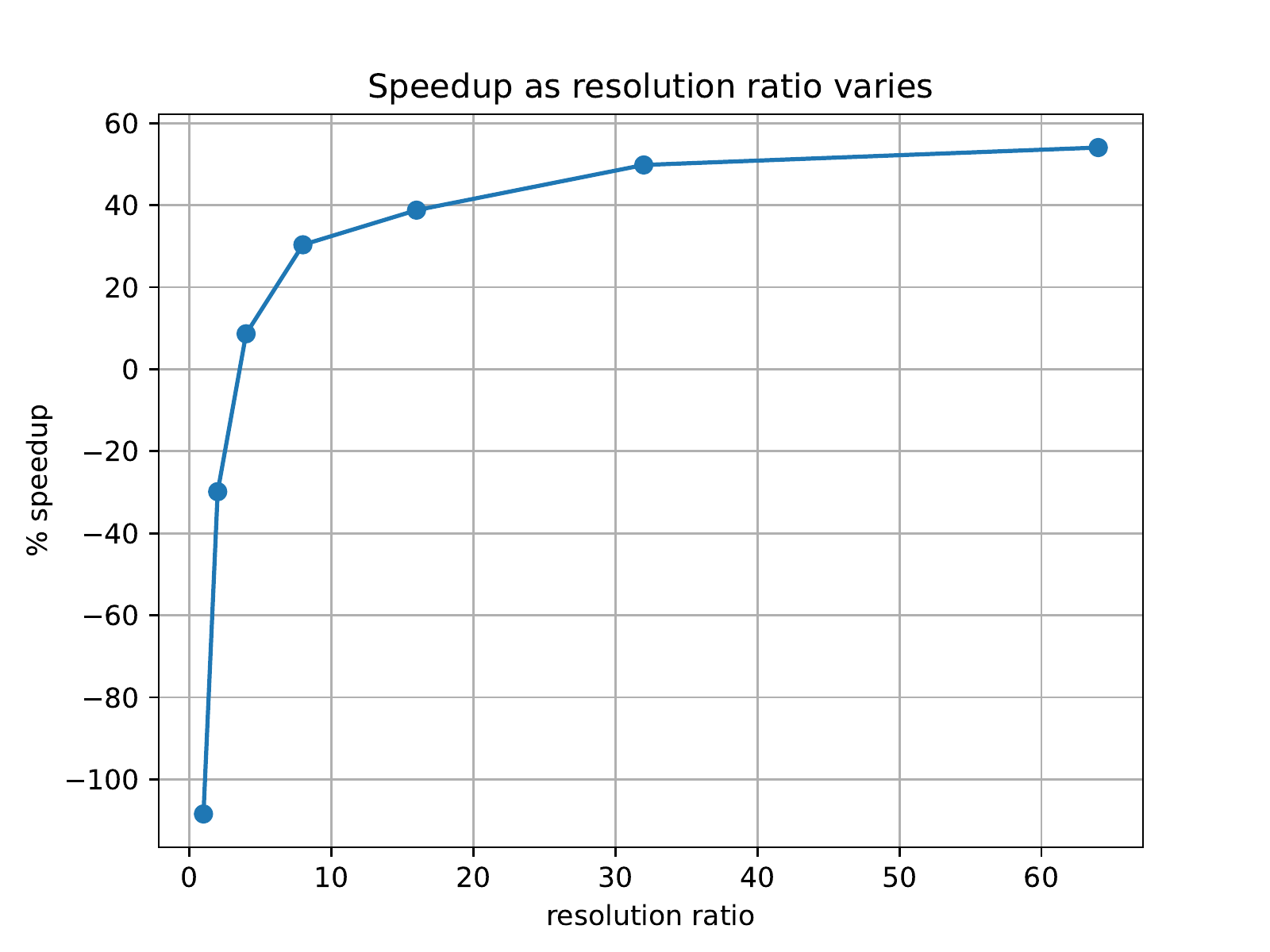}
        \caption{~}
        \label{fig:resolution_ratio}
    \end{subfigure}
    \caption{
        Speedup versus mesh design parameters: (a) varying count ratio while resolution ratio is held constant at 32; (b) varying resolution ratio as count ratio is held near one. In both cases, speedup increases and then levels off at 50--70\%. Speedup is calculated as in Eq. \ref{eqn:speedup}.
    }
    \label{fig:lts_mesh_parameters}
\end{figure}

In a preliminary analysis, we investigated how varying the count and resolution ratios affected the performance of LTS in an idealized test case.
Namely, in the shallow-water core of MPAS-O, we created a test case following test case five from \citeA{williamson1992} and generated a series of meshes that had varying values for the count ratio and the resolution ratio.
We produced the plots in Figure \ref{fig:lts_mesh_parameters}, which are relevant to the work done here for two reasons.
First, the shallow-water equations used to generate this data are identical to those used in our Hurricane Sandy test case, except for additional forcing terms used here.
That is, we can expect for the results shown in Figure \ref{fig:lts_mesh_parameters} to be similar to the results we would get with the hurricane Sandy test case.
Second, this data clearly demonstrates the importance of both the count and resolution ratios when considering the performance of LTS.
We can see that as either parameter increases, the speedup increases, and that this increase is drastic close to zero.
The curves in both plots eventually level off as the speedup becomes limited by the other parameter, which is fixed.

\begin{figure}[h!]
    \centering
    
    \includegraphics[width=0.95\textwidth]{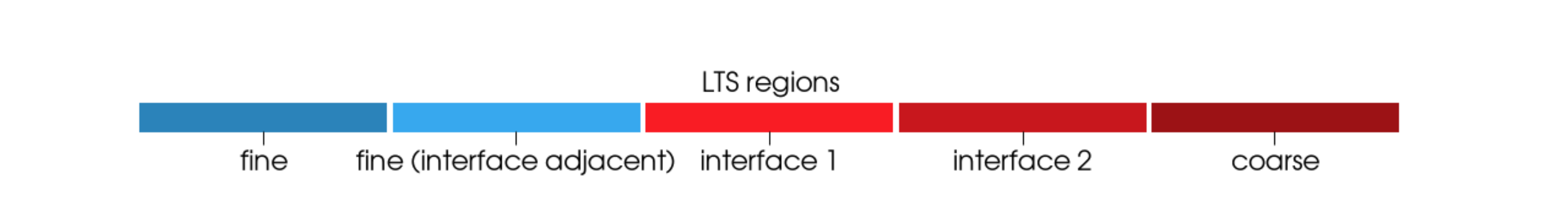}
    \vspace{-5pt}
    
    \begin{subfigure}{0.45\textwidth}
        \includegraphics[width=\textwidth]{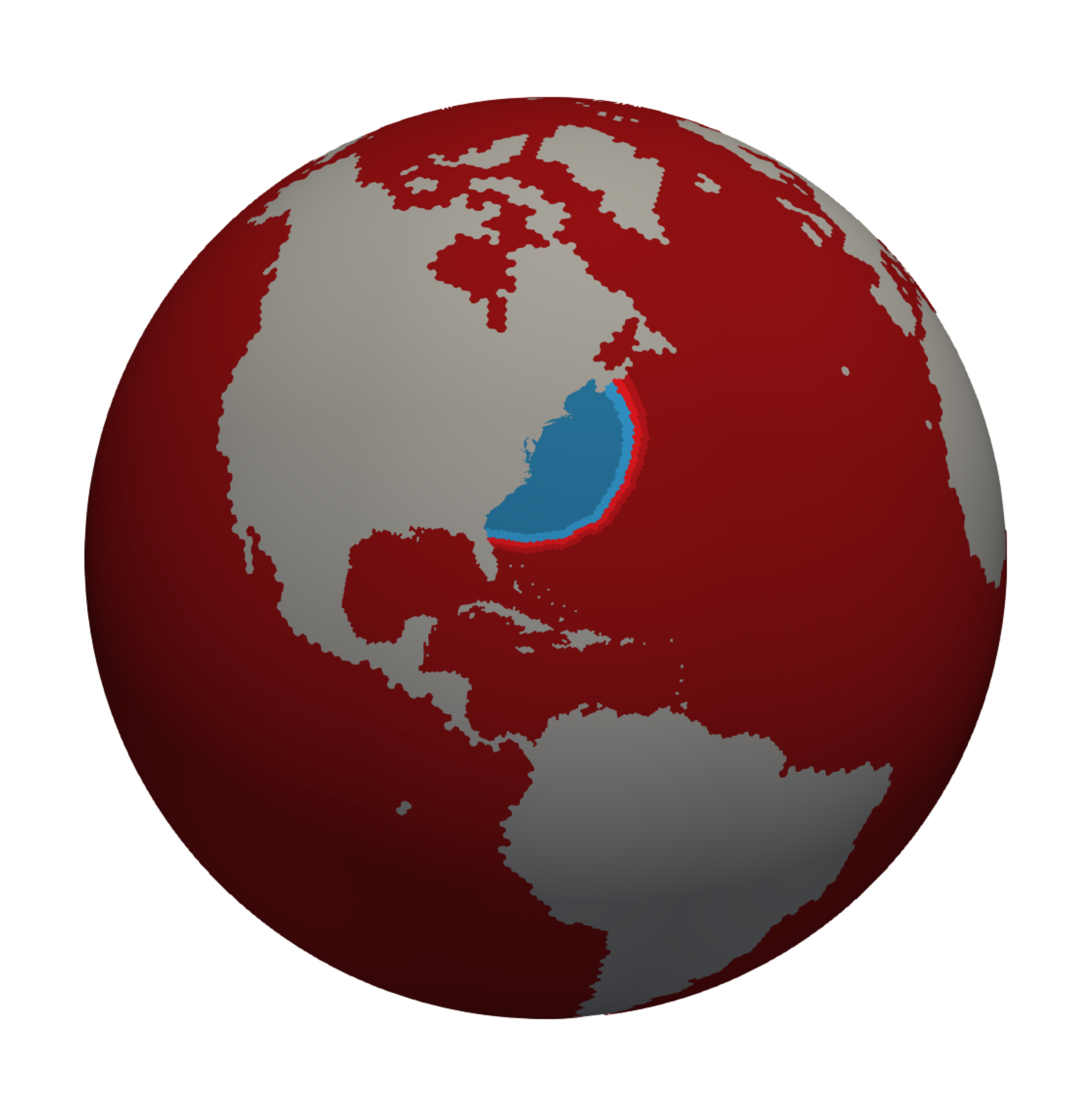}
        \caption{East US coast (EC) fine region}
        \label{fig:ec_fine_region}
    \end{subfigure}
    \begin{subfigure}{0.45\textwidth}
        \includegraphics[width=\textwidth]{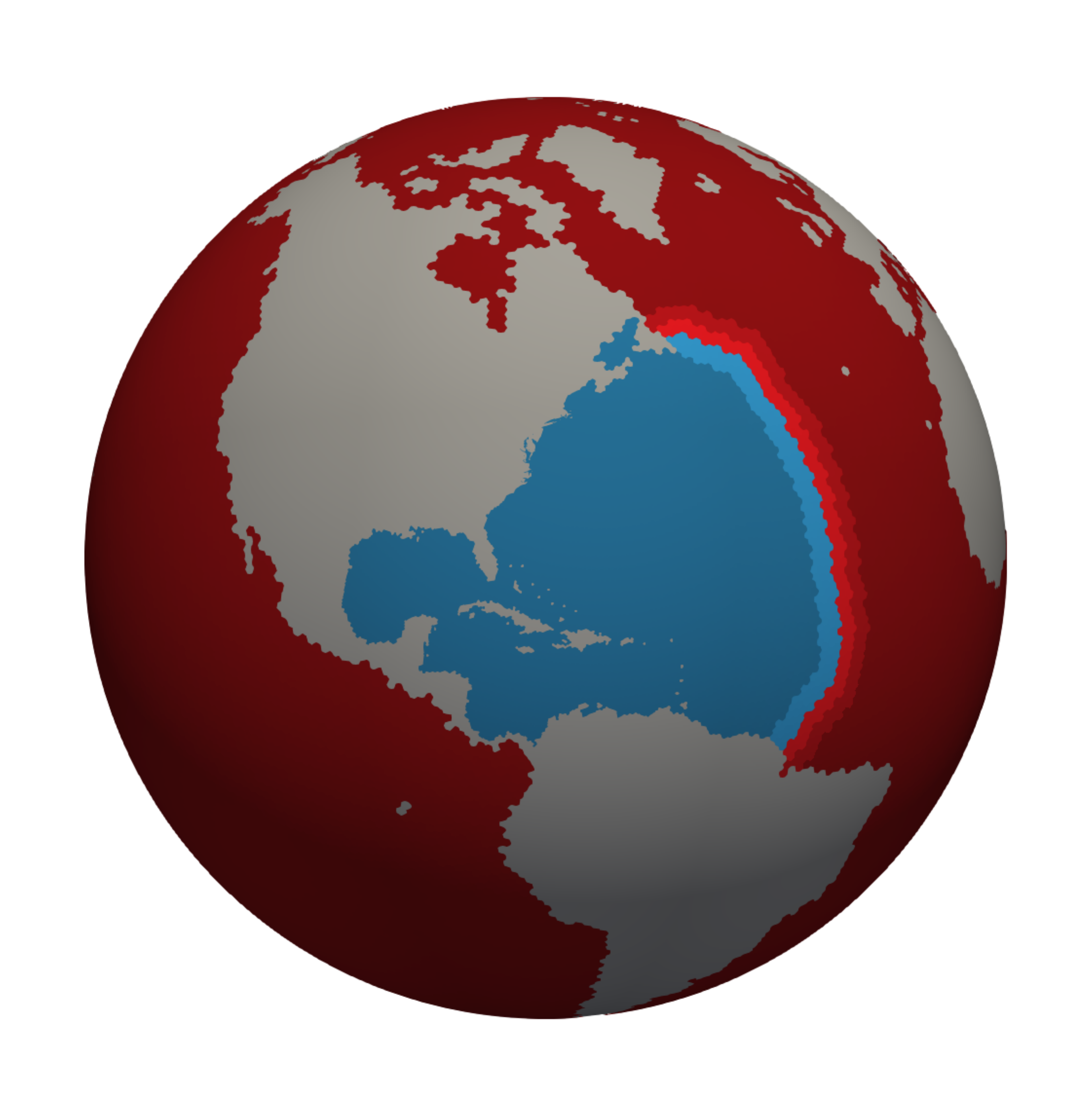}
        \caption{Western Atlantic (WA) fine region}
        \label{fig:wa_fine_region}
    \end{subfigure}
    \caption{The two configurations of the fine region used for DelBay meshes: the smaller East US Coast fine region (a) and the larger Western Atlantic fine region (b).}
    \label{fig:fine_regions}
\end{figure}

Figure \ref{fig:fine_regions} shows two configurations for the placement of the fine and coarse regions.
The configuration in Figure \ref{fig:ec_fine_region} will be referred to as the EC configuration and the configuration in Figure \ref{fig:wa_fine_region} will be referred to as the WA configuration.
We consider these different configurations because moving the fine region to include the western Atlantic cells greatly increases the size of our coarse time-step, with the penalty of adding a non-trivial number of cells to our fine region where we must use the fine time-step restricted by the smallest cell in the region.
This trade-off between the count ratio and the resolution ratio is a significant part of our investigation and will be discussed in Section \ref{subsec:performance}.

\subsubsection{Load Balancing, Communication, and Additional Interface Layers}
\label{subsubsec:load_balancing}

Performing the LTS3 algorithm in parallel requires careful consideration of load balancing.
LTS requires that cells in a given mesh be sorted into different groups, some are coarse cells, some are fine cells, and some are interface cells.
Each type of cell requires a different amount of work, so load balancing among MPI processes is non-trivial; an efficient way to address this has been a primary concern of \citeA{capodaglio2022}.
For effective load-balancing, each MPI rank is given three sets of cells, one set of fine cells, one set of coarse cells, and one set of interface cells.
The different groups of cells are spread across each rank so that each process has an approximately equal number of each type.
This helps to ensure that no process is idle while it waits for the others to finish work.
An issue with this scheme is that in most mesh configurations, there are very few interface cells relative to the number of fine or coarse cells.
As a result, when running with more than a few processes, there may not be enough interface cells in the mesh for an equal number to be spread across processes in a way that also respects the expense of MPI communication calls.
%
\begin{figure}[h!]
    \centering
    
    \includegraphics[width=0.95\textwidth]{graphics/meshes/ltsRegions_colorBar.pdf}
    \vspace{-5pt}
    
    \includegraphics[width=0.6\textwidth]{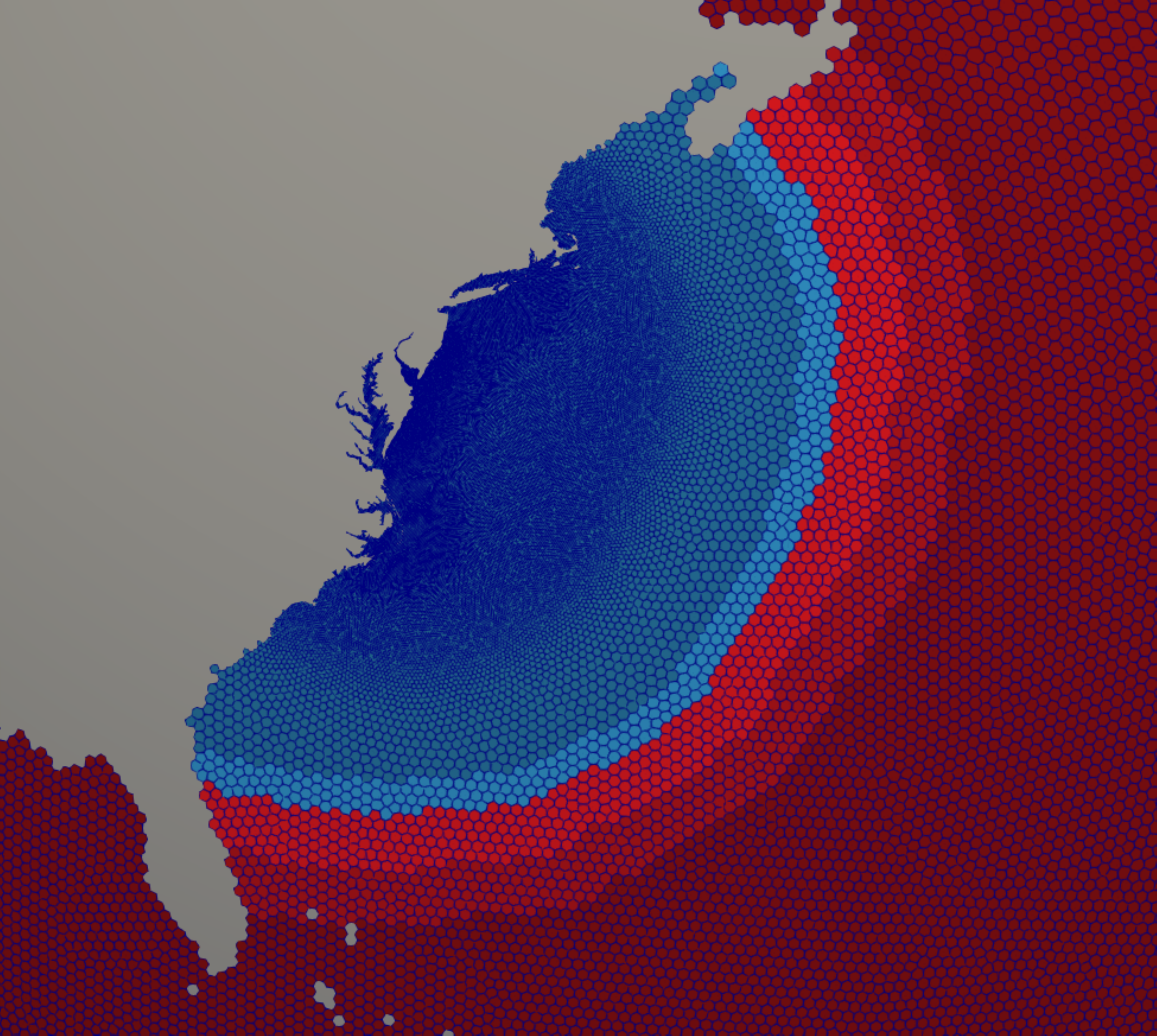}
    \caption{
        An example of a mesh needing additional layers of interface cells for load balancing in DelBay1km.
        Here, there are 10 interface layers.
    }
    \label{fig:tuned_interface_layers}
\end{figure}
To avoid this issue, additional interface layers are added to the mesh (see Figure \ref{fig:tuned_interface_layers}). Because LTS3 is \( \mathcal{O}\left( (\Delta t)^3 \right) \) everywhere on the mesh, these additional interface layers have no negative impact on the model's accuracy.
Adding these additional interface layers allows us to insure that there are enough interface cells to spread across a given number of processes.
It is shown in \citeA{capodaglio2022} that, as a rule of thumb, each process should own at least 100 interface cells to achieve sufficient load-balancing.
On each mesh and for each LTS fine region configuration, we have set a number of interface layers that ensures this requirement is met.
The number of interface layers used in each case is reported in Table \ref{tbl:meshes}.


\section{Results}
\label{sec:results}

In this section, we discuss the performance of LTS3 compared to RK4, in terms of computational time and accuracy of the sea-surface height predictions.


\subsection{Computational Time}
\label{subsec:performance}

\begin{figure}[h!]
    \centering
    \includegraphics[width=0.8\textwidth]{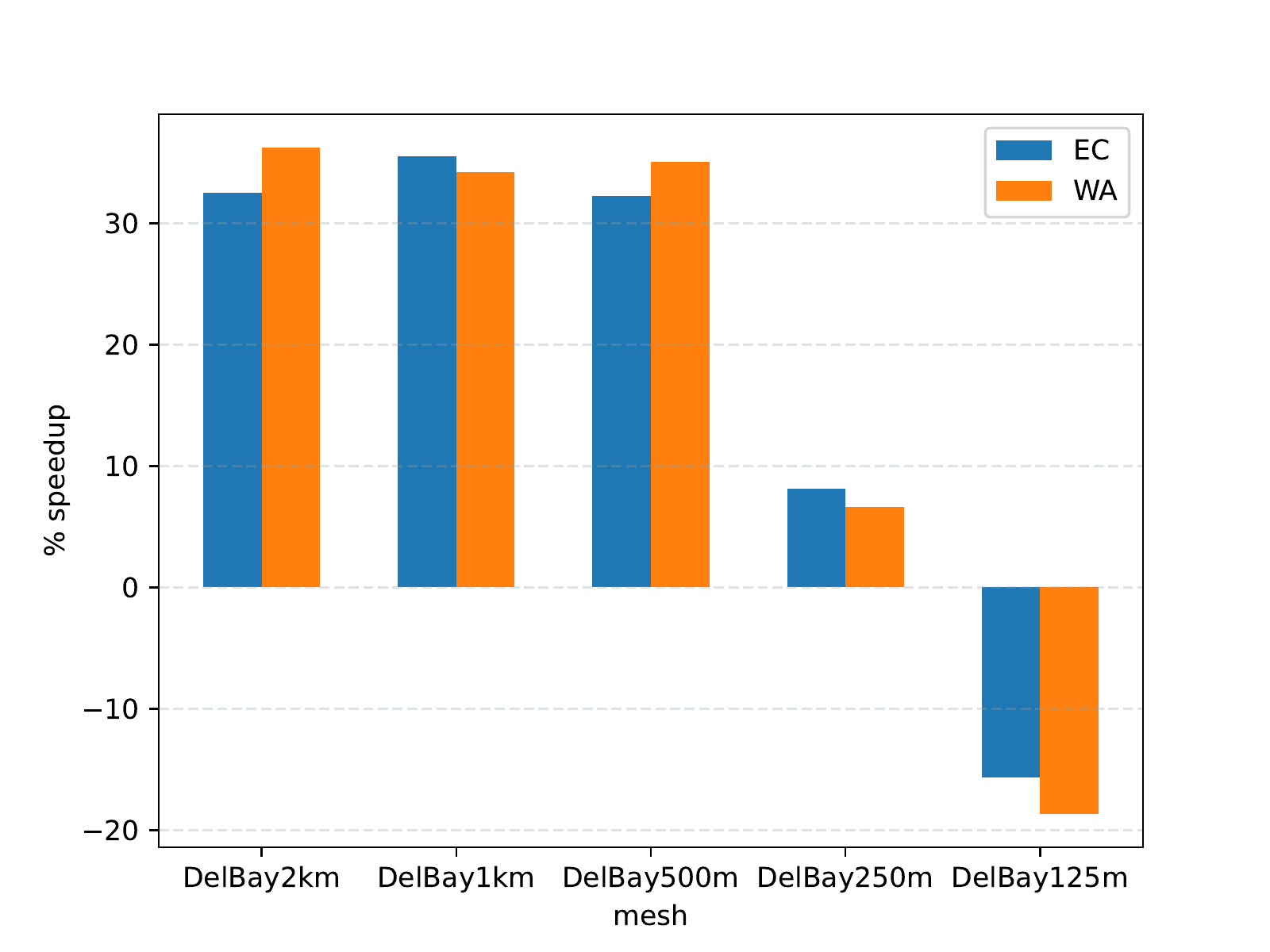}
    \caption{
        Speedup in terms of CPU-time obtained using LTS3 over RK4 in both the EC and WA configuration.
        Speedup is calculated as in Eq. \ref{eqn:speedup}.
    }
    \label{fig:performance}
\end{figure}

Figure \ref{fig:performance} reports the performance differences in terms of CPU-time of RK4 and LTS3 in both the EC and WA configurations.
To obtain this data, the model has been run using both RK4 and LTS3 for a set number of simulated seconds (referred to as the run-time) and the amount of time the model spends on time-stepping in both cases was recorded.
To account for differences in network communication speeds, timings are averaged over five simulations.
The speedup, or the percentage value that LTS3 is faster than RK4, is computed as
\begin{equation}
    \label{eqn:speedup}
    \mbox{speedup} = \frac{\text{time RK4} - \text{time LTS3}}{\text{time RK4}} \cdot 100. 
\end{equation}
All simulations were run on the Badger\footnote{
    Badger is a 660 node compute cluster where each node contains two 2.10GHz Intel Xeon E5-2695 v4 processors, each having 18 cores.
    Each node has 128GB of memory.
    Badger has a peak compute speed of 798 Tera-FLOPS per second.
}
cluster at Los Alamos National Laboratory.

Table \ref{tbl:performance} reports the specific results of each test case, including the values for the time integration timer in each case and the number of simulated seconds the model was run for.
One can note that, on each mesh, the run-time is different, and is selected to be the smallest number of seconds that is divisible by the coarse time-step used in the WA configuration, the coarse time-step used in the EC configuration, and the time-step for RK4.
This is to ensure that for each method and configuration, there is no case in which the model is actually advancing to a time further than intended, which would pollute the results.

\begin{table}[h!]
    \centering
    \resizebox{\columnwidth}{!}{%
        \begin{tabular}{lrrrrr}
            ~ & \textbf{DelBay2km} & \textbf{DelBay1km} & \textbf{DelBay500m} & \textbf{DelBay250m} & \textbf{DelBay125m} \\ \toprule
            run-time (s) & 6120 & 6840 & 960 & 960 & 960 \\ 
            number of MPI ranks & 2 & 8 & 32 & 64 & 178 \\ \midrule
            RK4 (s) & 57.92 & 133.26 & 47.57 & 95.76 & 208.58 \\ \bottomrule
            LTS3 EC (s) & 39.10 & 85.94 & 32.24 & 87.98 & 241.19 \\
            speedup &  32.50 & 35.51 & 32.23 & 8.13 & -15.63 \\ \bottomrule
            LTS3 WA (s) & 36.94 & 87.65 & 30.89 & 89.39 & 247.52 \\ 
            speedup & 36.23 & 34.23 & 35.05 & 6.65 & -18.67
        \end{tabular}
    }
    \caption{
        CPU-time performance of RK4 and LTS3 in both the EC and WA configuration.
        Speedup is calculated as in Eq. \ref{eqn:speedup}.
        The number of cells per process on each mesh is approximately between 25,000 and 29,000. 
    }
    \label{tbl:performance}
\end{table}

In another storm surge focused application of LTS schemes, \citeA{dawson2013} are able to cut the computational time of their model in half using LTS.
We observe that, while the non-LTS global scheme used for comparison in \citeA{dawson2013} is technically of a higher order than their LTS scheme, the smallest time-step used by said LTS scheme is the same as the global time-step used by the non-LTS scheme.
That is, the non-LTS scheme does not take advantage of the fact that as a higher order scheme, it admits a larger time-step.
As such the performance results from \citeA{dawson2013} are not directly comparable to our own.

It is well known that in parallel applications, communication between processors is more expensive than floating-point operations in terms of computational time, and that on a fixed mesh, past a certain number of processes, a time-stepping scheme will become dominated by communication between processors and will not scale efficiently.
Because LTS3 requires that the computational domain be decomposed into different classes of cells that require different work-loads, LTS3 requires more communication than RK4; as the number of processes increases, communication overhead will have a stronger impact on LTS3 than on RK4.
For a fair comparison between the two schemes, the runs from Table \ref{tbl:performance} were obtained with a number of processes for which communication did not become prevalent for either RK4 or LTS3.

\begin{table}[h!]
    \centering
    \resizebox{\columnwidth}{!}{%
        \begin{tabular}{lrrrrr}
            ~ & \textbf{DelBay2km} & \textbf{DelBay1km} & \textbf{DelBay500m} & \textbf{DelBay250m} & \textbf{DelBay125m} \\ \toprule
            run-time (s) & 6120 & 6840 & 960 & 960 & 960 \\ 
            number of MPI ranks & 16 & 64 & 256 & 512 & 1536 \\ \midrule
            RK4 (s) & 9.57 & 20.73 & 5.96 & 12.05 & 23.35 \\ \bottomrule
            LTS3 EC (s) & 8.49 & 17.93 & 5.90 & 38.51 & 83.90 \\
            speedup &  11.31 & 13.50 & 1.00 & -219.73 & -259.36 \\ \bottomrule
            LTS3 WA (s) & 7.75 & 17.71 & 5.29 & 26.19 & 90.60 \\ 
            speedup & 19.07 & 14.56 & 11.23 & -117.46 & -288.05
        \end{tabular}
    }
    \caption{
        CPU-time performance of RK4 and LTS3 at high process counts.
        The number of cells per process on each mesh is approximately 3,000.
    }
    \label{tbl:performance_high_proc_count}
\end{table}

For completeness, in Table \ref{tbl:performance_high_proc_count}, we give performance results at higher process counts where there are only 3,000 cells per process as opposed to more than 25,000 cells per process as in Figure \ref{fig:performance} and Table \ref{tbl:performance}.
Here, we see that LTS3 suffers from increased communication time and therefore speed-ups are less dramatic than in  Table \ref{tbl:performance}, although LTS3 still remains faster than RK4 even in this case, for all but one of the meshes it was faster on before.
The increased communication time in this case is due to each processor not having enough work to do during each time-step, because this is a single layer model.
In a layered model with \( n \) layers, there is approximately \( n \) times more work to be done per cell per time-step, and it has been shown in \citeA{capodaglio2022} that in a shallow water model with 100 layers, LTS can be run effectively with as little as 500 cells per process without communication becoming dominant.

From the results in Figure \ref{fig:performance}, there are a few things we learn about efficient mesh design to achieve optimal performance with LTS.
First, consider DelBay500m in the WA configuration where we see a speedup of 35\%.
As noted in Section \ref{subsubsec:lts_mesh_parameters}, we are particularly interested in the count ratio and resolution ratio of a mesh when considering the potential for good performance with LTS (see Figure \ref{fig:lts_mesh_parameters}).
In DelBay500m, the count ratio is 1.28 and the resolution ratio is 60, i.e. there are 1.28 times as many coarse cells than fine cells, and the smallest fine cells are 60 times smaller than the smallest coarse cells.
On the other hand, for the DelBay250m in the WA configuration the speed up is 6.65\%.
Here, the count ratio is only 0.39 (meaning that there are approximately 2.56 times more fine cells than coarse cells) and the resolution ratio is 120.

The DelBay250m WA case has the advantage over the DelBay500m WA case in having a higher resolution ratio, but in turn has a much lower count ratio.
This means that in the DelBay250m WA case, LTS3 spends the majority of its effort time-stepping on the fine region with a time-step that is smaller than the global time-step used by RK4, and even though the coarse time-step is much larger than RK4's global time-step, there are fewer cells where it is used.
This points to the fact that, when designing a mesh with the desire to take advantage of the benefits of LTS, one needs to take into account the count ratio so that the mesh is not overloaded with fine time-step cells.
In the case of DelBay125m in both the EC and WA configurations, when the count ratio is especially low (0.12 and 0.11 respectively), we see that LTS3 does not perform as well as RK4 in terms of CPU-time.
In extreme cases such as this where the mesh is predominantly composed of high-resolution cells, the use of an LTS scheme does not make sense, and one might benefit from simply using a higher-order global method that admits a larger time-step on the mesh's smallest cells.

Another important consideration is the placement of the border between the fine and coarse regions that defines which cells use the fine time-step and which use the coarse time-step.
In meshes with only two regions of differing resolutions there is no choice to make, but on meshes such as the ones used here, where there are more than two resolutions, there are multiple choices for the placement of the fine region.
What this choice should be is not always clear.
As can be seen in Figure \ref{fig:performance}, there are some cases where the EC configuration performs better and some where the WA configuration performs better.
In the case of DelBay250m and DelBay125m, the performance in both configurations is not particularly good.
However, that does not mean that LTS cannot perform well on these meshes.
The LTS configurations chosen in this work are experimental and there may be other configurations not investigated here that could be used to improve performance on these meshes.

In summary, LTS schemes require that the user consider both the count and resolution ratios as they configure a mesh for LTS, but this extra time spent is easily worth the greater than 35\% speedups that can be achieved when the configuration is done well.
As a general rule, one's goal is to maximize both of these parameters.
It is particularly important that the count ratio not be too low.
Figure \ref{fig:count_ratio} suggests that a count ratio of 0.2 (1:5) as a rule-of-thumb cut-off.
The results in Figures \ref{fig:resolution_ratio} and \ref{fig:performance} suggest that increasing the resolution ratio has diminishing returns if it is detrimental to the count ratio.
That is, if the vast majority of work is being done with the fine time-step, it does not matter if the coarse time-step is many times larger than the fine time-step.
To give some general advice, the quality of which will doubtlessly vary across applications, it is best to first achieve a reasonable value for the count ratio even at the expense of the resolution ratio as long as the resolution ratio can be kept reasonably high.
If this is not possible, on may wish to use a method that admits a larger time-step altogether like RK4; on a multi-resolution mesh predominately populated by small cells it does not make sense to use a LTS scheme.


\subsection{Accuracy}
\label{subsec:accuracy}

\begin{figure}[h!]
    \centering

    \begin{subfigure}{0.49\textwidth}
        \includegraphics[width=\textwidth]{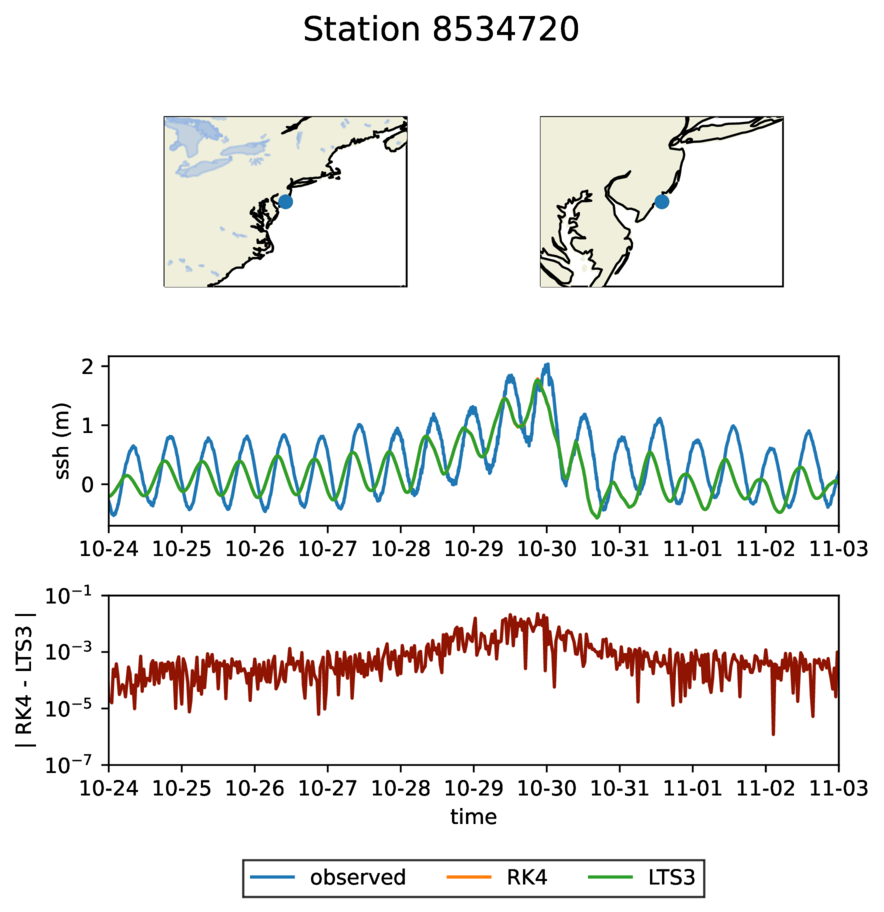}
        \caption{~}
        \label{fig:4_ec_diff_8534720}
    \end{subfigure} \hfill
    \begin{subfigure}{0.49\textwidth}
        \includegraphics[width=\textwidth]{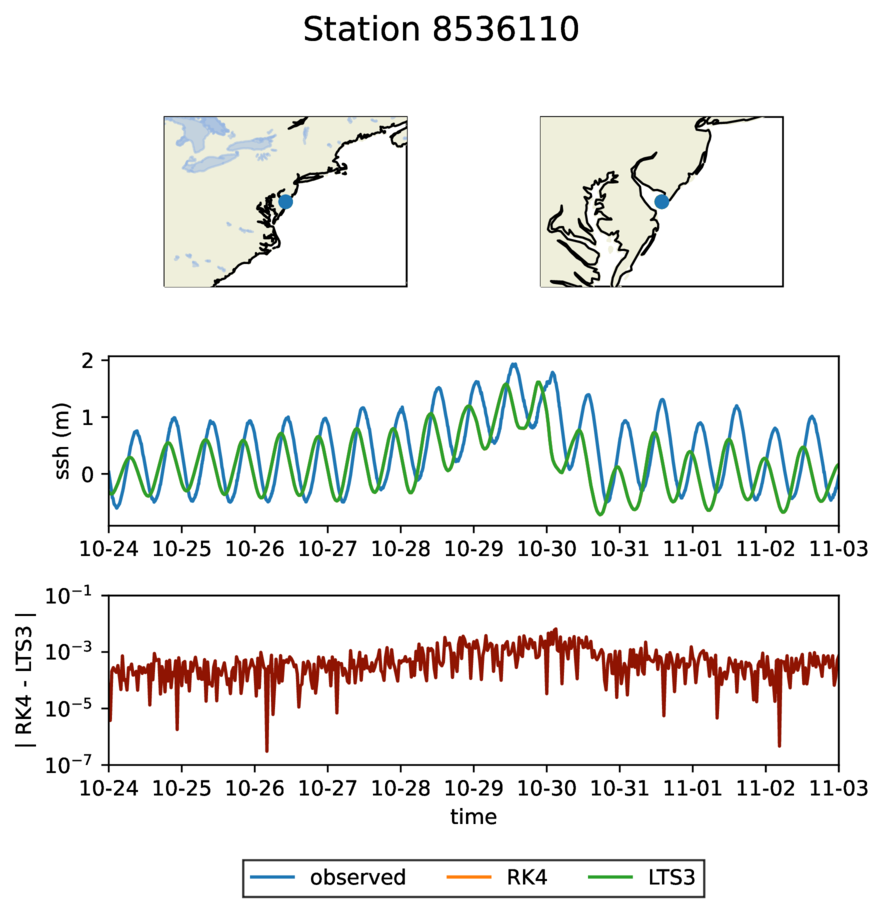}
        \caption{~}
        \label{fig:4_ec_diff_8536110}
    \end{subfigure} \\ \bigskip

    \begin{subfigure}{0.49\textwidth}
        \includegraphics[width=\textwidth]{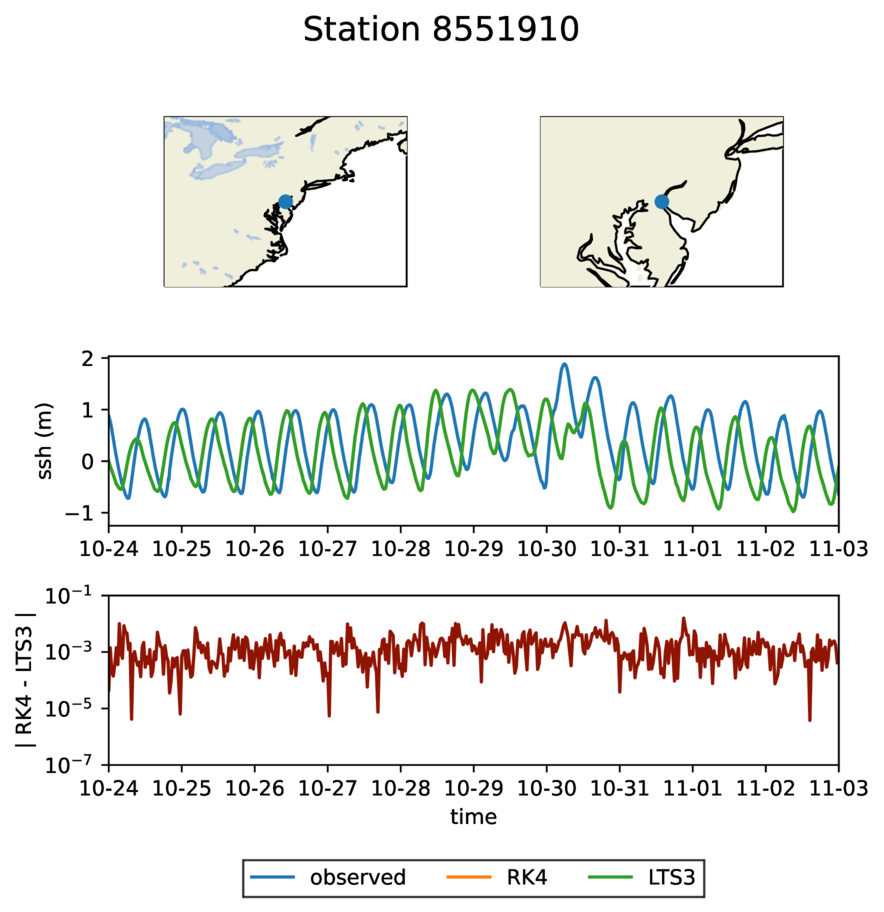}
        \caption{~}
        \label{fig:4_ec_diff_8551910}
    \end{subfigure} \hfill
    \begin{subfigure}{0.49\textwidth}
        \includegraphics[width=\textwidth]{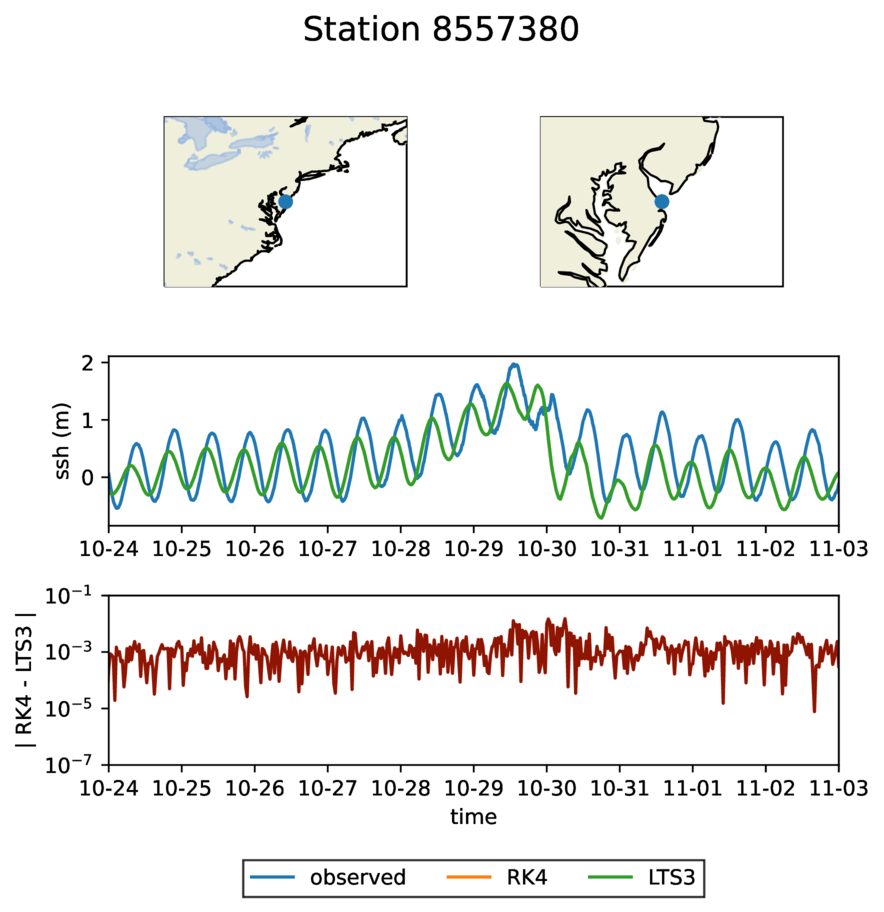}
        \caption{~}
        \label{fig:4_ec_diff_8557380}
    \end{subfigure}
    
    \caption{
        SSH solutions from RK4 and LTS3 compared to observed tide gauge data on DelBay125m.
        Note that the curve corresponding to LTS3 is covering the curve corresponding to RK4 in the upper plots. The absolute difference between the two schemes is shown on a log scale in the lower plots.
    }
    \label{fig:4_ec_diff}
\end{figure}

Here we compare the sea-surface height (SSH) predicted by RK4 and LTS3 to observed data.
The observed data are from NOAA's Center for Operational Oceanographic Products and Services (CO-OPS) gauges and are available at \url{https://tidesandcurrents.noaa.gov/}.
The SSH predictions produced by LTS3 and RK4 are nearly identical (Figure \ref{fig:4_ec_diff}); their difference is, at most, on the order of centimeters.

\begin{figure}[h!]
    \centering

    \begin{subfigure}{0.49\textwidth}
        \includegraphics[width=\textwidth]{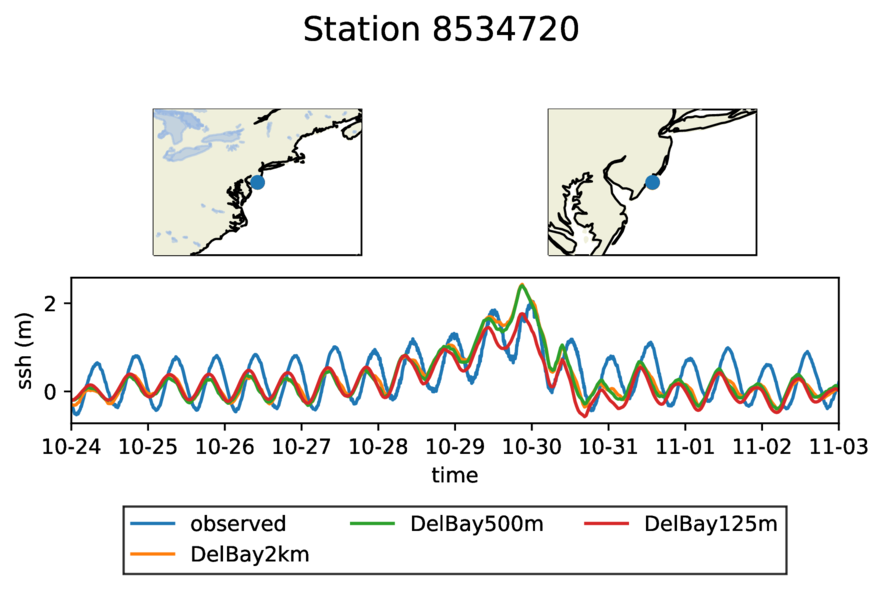}
        \caption{~}
        \label{fig:ec_lts_all_8534720}
    \end{subfigure} \hfill
    \begin{subfigure}{0.49\textwidth}
        \includegraphics[width=\textwidth]{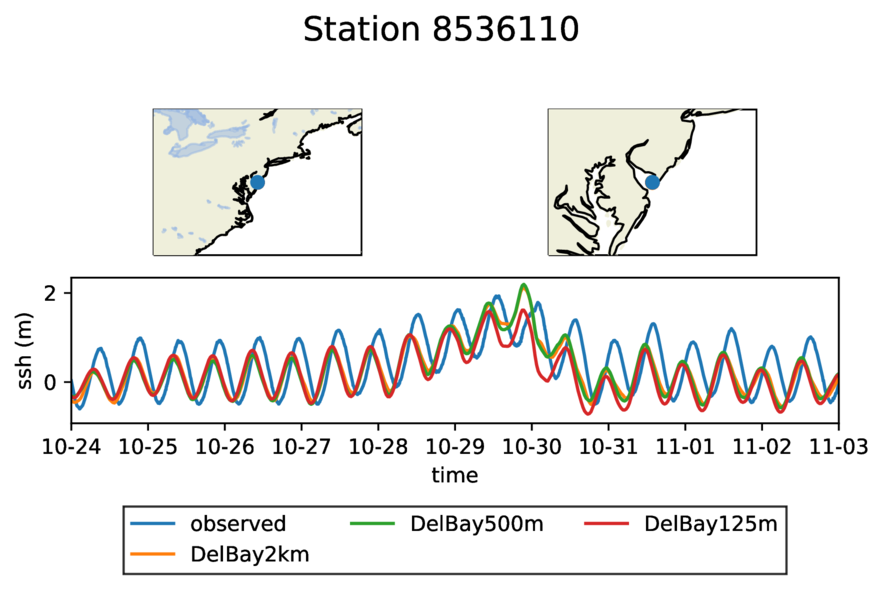}
        \caption{~}
        \label{fig:ec_lts_all_8536110}
    \end{subfigure} \\ \bigskip

    \begin{subfigure}{0.49\textwidth}
        \includegraphics[width=\textwidth]{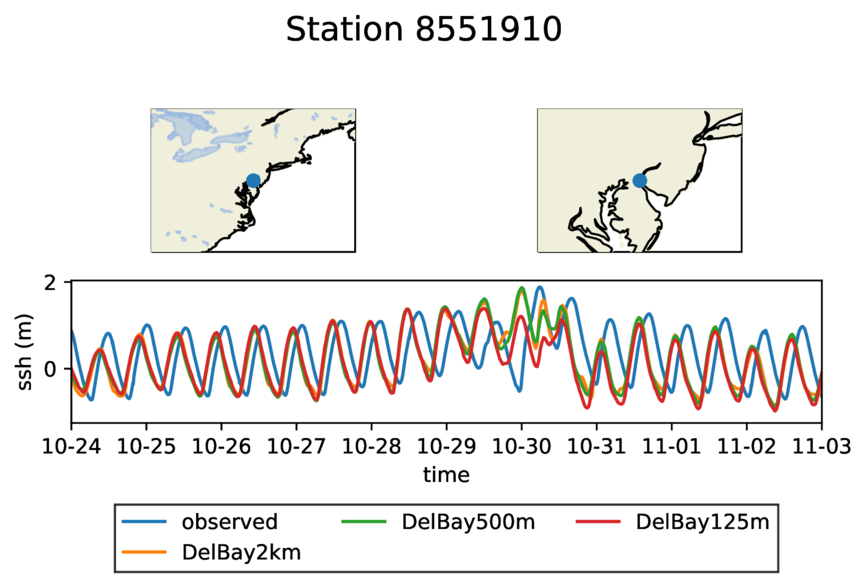}
        \caption{~}
        \label{fig:ec_lts_all_8551910}
    \end{subfigure} \hfill
    \begin{subfigure}{0.49\textwidth}
        \includegraphics[width=\textwidth]{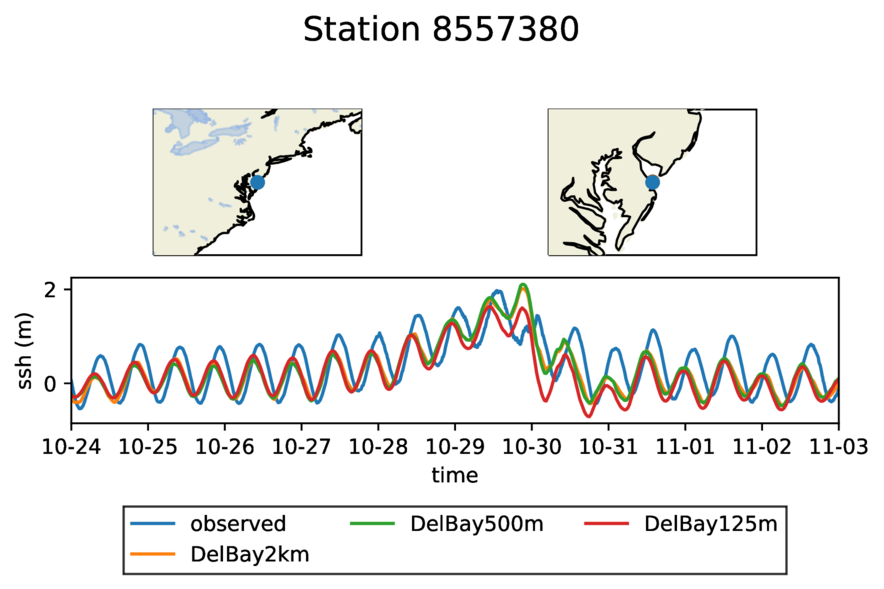}
        \caption{ }
        \label{fig:ec_lts_all_8557380}
    \end{subfigure}
    
    \caption{Comparison of LTS3 SSH solutions on DelBay2km, DelBay500m, and DelBay125m.}
    \label{fig:ec_lts_all}
\end{figure}

The SSH solutions produced by LTS3 on DelBay2km, DelBay500m, and DelBay125m do not differ greatly in terms of accuracy compared to observed data (Figure \ref{fig:ec_lts_all}).
At present, the use of a finer mesh on the coast does not greatly improve the SSH prediction at tidal gauges, meaning that the model itself needs improvement and that any deficiency in the solutions are is due to LTS3.
The purpose of this work is not necessarily to produce an accurate storm-surge model, but rather to show that LTS schemes, particularly LTS3, can produce results comparable to RK4 in a shorter amount of time.
Our results support this claim.

\section{Conclusion}
\label{sec:conclusion}

We have used a third order LTS scheme (LTS3) to model the storm surge of Hurricane Sandy in the Delaware Bay, using meshes of unprecedentedly high resolution for MPAS-O, and have shown that the solutions obtained are qualitatively comparable to those produced by the classical four stage, fourth order Runge-Kutta scheme (RK4).
Furthermore, LTS3 produces these solutions in considerably less computational time than RK4, with speedups of up to 35\%.
This was the first real-world scientific application of LTS schemes in the framework of MPAS-O, and will be used to pave the way for further use of LTS within the MPAS framework for the accurate capture of coastal physical phenomena requiring high spatial resolution, without compromising the overall speed of the simulation.
In the future, we are interested in exploring the deployment of LTS as a tool to enhance existing time-stepping schemes in MPAS-O such as the split-explicit scheme, where a model similar to the one considered in this work could be solved with LTS during the barotropic step. Work is also underway to use LT3 within a multi-layer primitive ocean model, feature that would be useful to employ LTS also in the baroclinic mode of the split-explicit solver.


\section*{Open Research}

The source code for the LTS development branch of MPAS-O used here can be found on GitHub and Zenodo.\nocite{sourceCode2022}
\begin{itemize}[leftmargin=2.5cm]
    \item[GitHub:] \url{https://github.com/jeremy-lilly/MPAS-Model/tree/} \\ \url{4e1f5a3cfe78ef01afa07e61f0d46670fdb6c014}
    \item[Zenodo:] \url{https://doi.org/10.5281/zenodo.6904061} \\ DOI: \texttt{10.5281/zenodo.6904061}
\end{itemize}
The data generated for this paper are also publicly available on Zenodo.
This includes the model output used to generate SSH plots and the log files from performance experiments.\nocite{data2022}
\begin{itemize}[leftmargin=2.5cm]
    \item[Zenodo:] \url{https://doi.org/10.5281/zenodo.6908349} \\ DOI: \texttt{10.5281/zenodo.6908349} 
\end{itemize}

\acknowledgments
JRL was supported by the U.S. Department of Energy (DOE), Office of Science, Office of Workforce Development for Teachers and Scientists, Office of Science Graduate Student Research (SCGSR) program.
The SCGSR program is administered by the Oak Ridge Institute for Science and Education (ORISE) for the DOE.
ORISE is managed by ORAU under contract number DE‐SC0014664.
GC, MRP, SRB, and DE were supported by the Earth System Model Development program area of the U.S. DOE, Office of Science, Office of Biological and Environmental Research as part of the multi-program, collaborative Integrated Coastal Modeling (ICoM) project. This research used resources provided by the Los Alamos National Laboratory Institutional Computing Program, which is supported by the U.S. DOE National Nuclear Security Administration under Contract No. 89233218CNA000001.

\bibliography{references}

\end{document}